\documentclass[aps,prd,preprint,a4paper,showpacs,nofootinbib,
superscriptaddress]{revtex4-1}
\usepackage{bm}
\usepackage{indentfirst}
\usepackage{amsmath}
\usepackage{graphicx}
\usepackage{float}
\usepackage{amssymb}
\usepackage{subfigure}
\usepackage{amssymb}
\usepackage{hyperref}
\usepackage{epstopdf}
\usepackage[section]{placeins}
\usepackage[utf8]{inputenc}
\hypersetup{
    colorlinks=true,
    linkcolor=red,
    citecolor=blue,
}
\usepackage{color}

\begin{document}

\title{Exploring non-singular black holes in gravitational perturbations }
\author{Hang Liu}
\affiliation{School of Physics and Astronomy,
Shanghai Jiao Tong University, Shanghai 200240, China}
\affiliation{School of Aeronautics and Astronautics, Shanghai Jiao Tong University, Shanghai 200240, China}
\author{Chao Zhang}
\affiliation{School of Physics, Huazhong University of Science and Technology, Wuhan, Hubei 430074, China}
\author{Yungui Gong}
\affiliation{School of Physics, Huazhong University of Science and Technology, Wuhan, Hubei 430074, China}
\author{Bin Wang}
\email{Corresponding Author: Bin Wang, wang\_b@sjtu.edu.cn}
\affiliation{Center for Gravitation and Cosmology, Yangzhou University, Yangzhou 225009, China.}
\affiliation{School of Aeronautics and Astronautics, Shanghai Jiao Tong University, Shanghai 200240, China}
\author{Anzhong Wang}
\affiliation{GCAP-CASPER, Physics Department, Baylor University, Waco, TX, 76798-7316, USA}

\begin{abstract}
We calculate gravitational perturbation quasinormal modes (QNMs) of non-singular Bardeen black holes (BHs)
and singularity-free BHs  in conformal gravity and examine their spectra in
wave dynamics comparing with standard  BHs in general relativity. After testing the validity
of the approximate signal-to-noise ratio (SNR) calculation for different space based interferometers,
we discuss the SNR of non-singular  BHs in single-mode gravitational waveform detections in LISA, TianQin and TaiJi.
We explore the impact of the Bardeen parameter and the conformal factor on the behavior of the SNR and find that
in comparison with the standard Schwarzschild  BHs, the increase of non-singular parameters leads to
higher SNR for more massive non-singular BHs. We also examine the effect of the galactic
confusion noise on the SNR and find that a dip appears in SNR due to such noise.
For non-singular BHs, with the increase of the non-singular parameters
the dip emerges for more massive BHs. It suggests  a wider range of  mass of non-singular black holes whose SNR will not be lowered by galactic noise, which implies the SNR for the super massive black holes centered at our galaxy will less likely to be influenced by galactic noise in non-singular black holes models than singular case in general relativity. We conduct  Fisher analysis which suggests that the non-singular black holes parameters can be detected accurately with measurement errors as small as $\sim10^{-4}-10^{-5}$.
The detections of non-singular  BHs are expected to be realized more likely by LISA or TaiJi.
\end{abstract}

\maketitle

\section{Introduction}

Gravitational waves (GWs) were predicted by Einstein's general relativity (GR)
a century ago and since then huge amount of effort has been put
into investigating the existence of GWs and thereby testing GR. Recently,
the first  GW event GW150914 with relatively high network SNR \cite{LIGOScientific:2018mvr} was detected by
LIGO Scientific Collaboration and Virgo Collaboration \cite{Abbott1}.
More detections of GWs \cite{LIGOScientific:2018mvr,Abbott:2020niy} have been reported afterwards.
The landmark detection
of GWs  ended the era of a single electromagnetic wave channel observation of our universe.
Successful detections of GWs
announced the dawn of multi-messenger astronomy with GW as a new probe to study our universe. As a new probe, people have proposed applications of GWs in the study of diversity subjects, such as searching dark matter \cite{Michimura:2020vxn,DeRocco:2018jwe,Obata:2018vvr,Liu:2018icu,Nagano:2019rbw,Martynov:2019azm,Grote:2019uvn,Morisaki:2018htj,Pierce:2018xmy,Manley:2020mjq},
exploring dark energy \cite{Weiner:2020sxn,Garoffolo:2020vtd,Singh:2020nna,Noller:2020afd,Yang:2020wby,Li:2019ajo,Zhang:2019ple}, cosmological parameter estimation with GW standard siren \cite{Jin:2020hmc,Zhao:2019gyk,Zhang:2019mdf,Zhang:2019loq,Wang:2019tto,Zhang:2019ylr}, searching for modified theories of gravity \cite{Nunes:2020rmr,Mastrogiovanni:2020gua,Niu:2019ywx,Ma:2019rei}, etc.

The ground based
interferometers (such as LIGO and Virgo) inevitably suffer from the gravity
gradient noise and seismic noise which lead to the limitation
that the detection of GWs with frequencies lower than 10Hz
is extremely challenging. 
However, it is of great
significance to probe GWs in lower frequency bands because a large number of GW sources containing rich physics are expected to fall in the
frequency bands from millihertz to hertz \cite{Kormendy}.
Among these sources, the mergers of massive black hole (MBH) binaries with mass
between $10^3M_{\odot}$ and $10^7M_{\odot}$ are expected
to happen frequently \cite{Hu2017,Barausse:2014oca,Klein:2015hvg} with total number up to $\sim 332$ of mergers per
year \cite{Klein:2015hvg},
although there is no conclusive evidence yet.
The existence of MBHs has been
confirmed in the center of galaxies, for example a black hole
named Sagittarius A* with mass about $4\times 10^6M_{\odot}$ was
discovered in the center of our Milky Way \cite{Abuter:2018drb}.
To detect GWs from intermediate and super massive sources, we have to move
our detectors to space. Laser Interferometer Space
Antenna (LISA) \cite{amaroseoane2017laser}, TaiJi \cite{Hu:2017mde}
and TianQin \cite{Luo} are space based detectors to probe GWs with frequencies
in the millihertz to hertz band.

Scientifically detecting GWs in space can disclose more physics of gravity. For the compact binaries,
LIGO and Virgo detected GW emission from the merger of binary neutron stars (BNS) (e.g. GW170817 \cite{Abbott7}).
The characteristic of BNS was confirmed and the BNS was distinguished from the merger of binary
black holes (BBHs) with the help of the electromagnetic observation
(e.g. gamma-ray burst \cite{Monitor:2017mdv,Goldstein:2017mmi} and
kilonova \cite{Arcavi:2017xiz,Coulter:2017wya,Lipunov:2017dwd,Soares-Santos:2017lru,Tanvir:2017pws,Valenti:2017ngx}).
Good sky location of the source can help for the detection of electromagnetic signals, just as in the case of event GW170817 \cite{Abbott7}.
However the electromagnetic signal of BNS might not be  detectable in some situations. One main reason is due to the large distance to the source and poor sky location which made the detection of electromagnetic signal quite difficult \cite{Abbott:2020uma,Chen:2020fzm}. Another reason might be
that the initial BNS are massive enough and will directly collapse into black holes  (BHs) \cite{Chen:2020fzm,Shibata:2019wef,Coughlin:2018fis,Kiuchi:2019lls} leaving negligible matter outside, then very faint electromagnetic signal would be created.
It is expected that space based GW detectors can help uncover the tidal deformability in the binary system,
whether it is zero or not \cite{Flanagan:2007ix} can serve to determine the binary system being BBH or BNS.
This signature can help understand better the galactic compact binaries.
In addition, the quantum effect arising from the quantum correction to classical gravity theory
around black hole event horizons can be encoded in GWs.
It was argued that such quantum effect can be detected through the observation of GW echoes \cite{Cardoso:2016rao,Cardoso:2016oxy}.
A tentative evidence for the echoes in GWs  was disclosed in the LIGO observations \cite{Abedi:2016hgu}, whereas it was claimed in Ref. \cite{Westerweck:2017hus,Tsang:2019zra} that current observations only provide low statistical significance for the existence of echoes. Nevertheless, it is expected that we can detect signals of echoes by the future space-based detectors or put strong constraints on alternative sources.
The waveforms of GWs detected by ground-based detectors \cite{Abbott1,Abbott2,Abbott3,Abbott4,Abbott5,Abbott6}
successfully confirmed GR in the  nonlinear and strong-field regimes.
However, we know that GR is not
complete, and in particular it is plugged with the singularity problem, the non-renormalization problem and has difficulties in understanding the universe at very large scales. These
provide
the motivations for conceiving modified theories of gravity. Whether the meddling with GR can produce GWs to be detected by ground based or space based detectors is an interesting question to be studied.

In this work we will concentrate on the study of an alternative theory of GR to accommodate singularity free black hole solutions to avoid the singularity problem in GR. The existence of singularities in the solutions to Einstein's field equations
has been a longstanding problem.
An idea considering the quantum effect
of gravity was proposed to eliminate singularities. Following this idea some
attempts have been made \cite{Ashtekar:2005qt,Nicolini:2005vd,LopezDominguez:2006wd,Hossenfelder:2009fc,Bojowald:2018xxu,Ashtekar:2018lag,Ashtekar:2018cay,Bodendorfer:2019cyv,Ashtekar:2020ifw,Jusufi:2019caq,BenAchour:2018khr,BenAchour:2020bdt,BenAchour:2020mgu} to alleviate the singularity problem  although a consistent
quantum theory of gravity is still absent so far.
To cure the shortcoming of GR at infrared and
ultraviolet scales, a wide class of modified theories of gravity has
been  constructed  with the purpose of  addressing conceptual and experimental
problems emerged in  the fundamental physics and providing  at least
an effective description of quantum gravity \cite{Capozziello:2011et}.
Considering the non-physical characteristics of  singularities,
it is natural to find non-singular solutions to Einstein's equations.
The first non-singular black hole
solution was found by Bardeen and it was later revealed that the nonlinear
electromagnetic energy-momentum tensor playing the role of the source
term in field equations \cite{AyonBeato:1998ub}.
In the conformal gravity frame, the black hole singularity could be
removed under conformal transformations  by taking advantage
of the conformal symmetry of the spacetime \cite{Englert:1976ep,tHooft:2011aa,Dabrowski:2008kx,Mannheim:2011ds,Mannheim:2016lnx,Modesto:2016max,Bambi:2016wdn}.

It is of great interest to study the wave dynamics of such non-singular BHs and distinguish them from black hole solutions in GR. The study of wave dynamics outside BHs has been an intriguing subject for the last few decades (for recent review, see for example \cite{Konoplya:2011qq}).
A static observer outside a black hole can indicate successive stages of the wave evolution. After the initial pulse, the gravitational field outside the black hole experiences a quasinormal ringing, which describes the damped oscillations under perturbations in the surrounding geometry of a black hole with frequencies and damping times of the oscillations entirely fixed by the black hole parameters. The quasinormal mode (QNM) is believed as a unique fingerprint to directly identify the black hole existence
and distinguish different black hole solutions. We will employ the 13-th order WKB method
with averaging of the Pade approximations suggested first in
\cite{Matyjasek:2017psv} to compute the QNM of non-singular  BHs and compare it with the result of
wave dynamics in usual  BHs in GR. Since the numerical method we apply here has very high accuracy \cite{Konoplya:2019hlu}, we expect to find the special signatures of non-singular  BHs in the wave dynamics.

The detection of  QNMs can be realized through gravitational wave observations.
From the observational point of view, we can calculate the signal-to-noise ratio (SNR) from the
ringdown signals of GWs originated from the gravitational perturbations around  BHs.
Thus based on the precise QNM spectrum, we can obtain the SNR in GW  observations.
Different imprints in  QNMs caused by different black hole configurations can be reflected in behaviors of the SNR.  In order to distinguish different black hole solutions through the study of black hole spectroscopy,
we require large SNR in the black hole ringdown phase.
It was pointed out in \cite{Cardoso,Berti:2009kk} that to
resolve either the frequencies or damping time of fundamental
mode $(n=0)$ from the first overtone $(n^\prime=1)$ with the same
angular dependence $(l=l^\prime,m=m^\prime)$, the critical
value $\rho_{cri}$ of SNR  is required  to be around $\rho_{cri}\simeq100$,
while to resolve both the frequencies and damping time typically
requires $\rho_{cri}\simeq1000$.
The large SNR can serve as a smoking gun in GW  observations to identify the existence of non-singular black hole solutions in alternative theories of gravity. In the following discussion, we will not only examine the SNR in LISA, but  also extend the calculation of SNR to other space based GW observations, such as TaiJi and
TianQin, to check the feasibility of testing the existence of
non-singular  BHs.

The organization of the paper is as follows. In Section \ref{section2},
we introduce the calculation
of SNR for single-mode waveform  detections.  In Section \ref{section3},
we calculate the QNMs and the SNR for
non-singular BHs in conformal gravity.
In Section \ref{section4}, we generalize  such calculations to the case
for non-singular Bardeen  BHs. In Section \ref{section5}, we calculate the errors in parameter estimation through a
Fisher analysis.
Finally in the last section we present our  main conclusions.
In Appendix A, we prove that the approximate formula
in the SNR calculation developed in the context
of LISA is general and can be applied to TaiJi and TianQin observations within acceptable errors.

\section{The SNR for single-mode waveform}
\label{section2}

In this section, we give a brief review on how to calculate the SNR for a single-mode wave detection.
The basic idea was proposed in \cite{Cardoso} for LISA and we generalize the method to discuss the SNR for Tianqin and TaiJi. We should point out that following analysis is only applicable to the ringdown stage of the gravitational waves.

The gravitational waveform composed of cross component $h_{\times}$ and plus component $h_{+}$ emitting from a perturbed black hole (or from the distorted final black hole merging from supermassive black hole pairs) can be expressed as
\begin{equation}
h_+ + ih_{\times}=-\frac{2}{r^4}\int_{-\infty}^{+\infty}\frac{d\omega}{\omega^2}e^{i\omega t}\sum_{lm}S_{lm}(\iota,\beta)R_{lm\omega}(r),
\end{equation}
where $R_{lm\omega}(r)$ is the radial Teukolsky function \cite{Teukolsky} with the
approximation $R_{lm\omega}(r)\sim r^3 e^{-i\omega r} Z^{out}_{lm\omega}$ when $r\rightarrow\infty$. $Z_{lm\omega}^{out}$ is a
complex amplitude. Now we assume that the gravitational waveform can be written as a formal  QNM expansion
and  consider that the QNMs of  the Schwarzschild and  Kerr  BHs always exist in pairs (because QNMs with positive number $m$ and negative $-m$ exist at the same time, we denote real and imaginary part of  QNMs frequency with positive $m$ as $\{\omega_{lmn}, \tau_{lmn}\}$, and for negative $-m$ as $\{\omega'_{lmn},\tau'_{lmn}\}$) which should be included in the waveform expansion. In this way we have
\begin{equation}
\begin{split}
h_+ + ih_{\times}&=\frac{1}{r}\sum_{lmn}\left\{e^{i\omega_{lmn}t}e^{-t/\tau_{lmn}}S_{lmn}(\iota,\beta)Z^{out}_{lmn}+e^{i\omega_{lmn}^{\prime}t}e^{-t/\tau_{lmn}^{\prime}}S_{lmn}^{\prime}(\iota,\beta)Z^{\prime out}_{lmn}\right\}\\
&=\frac{M}{r}\sum_{lmn}\left\{\mathcal{A}_{lmn}e^{i(\omega_{lmn}t+\phi_{lmn})}e^{-t/\tau_{lmn}}S_{lmn}(\iota,\beta)+\right.\\
&\left.\qquad\qquad\qquad\qquad\mathcal{A}^{\prime}_{lmn}e^{i(-\omega_{lmn}t+\phi_{lmn}^{\prime})}e^{-t/\tau_{lmn}}S_{lmn}^{\ast}(\iota,\beta)\right\},
\end{split}
\end{equation}
where we have rewritten the complex $Z^{out}_{lmn}$ in terms of a real amplitude
$\mathcal{A}_{lmn}$ and a real phase $\phi_{lmn}$, and we factor out the black hole
mass $M$ by $Z^{out}_{lmn}=M\mathcal{A}_{lmn}e^{i\phi_{lmn}}$. In the above
expansion, $S_{lm}(\iota,\beta)$ stands for spin weighted spheroidal harmonics
whose complex conjugate is denoted by $S_{lmn}^{\ast}(\iota,\beta)$, $\iota$ and
$\beta$ are angular variables, and $l$, $m$ are indices analogous to those for
standard spherical harmonics corresponding to a particular case of
$S_{lmn}(\iota,\beta)$ in which both the perturbation field and black hole spin
are zero, $n$ denotes the overtone number. Note that we have the complex QNM
frequency $\omega=\omega_{lmn}+i/\tau_{lmn}$, where the real part denotes the
 oscillation frequency $\omega_{lmn}=2\pi f_{lmn}$ and the imaginary part
 $\tau_{lmn}$ is  the damping time of the perturbation oscillation.
For a single given mode labeled by ($l,m,n$), the real waveform measured
at the detector can be expressed as a linear superposition
of $h_+$ and $h_{\times}$
\begin{subequations}
\begin{align}
h_+=\frac{M}{r}\mathfrak{R}\left[\mathcal{A}_{lmn}^+e^{i(\omega_{lmn}t+\phi_{lmn}^+)}e^{-t/\tau_{lmn}}S_{lmn}(\iota,\beta)\right],\\
h_{\times}=\frac{M}{r}\mathfrak{I}\left[\mathcal{A}_{lmn}^{\times}e^{i(\omega_{lmn}t+\phi_{lmn}^{\times})}e^{-t/\tau_{lmn}}S_{lmn}(\iota,\beta)\right],
\end{align}
\end{subequations}
in which we have the relation $\mathcal{A}_{lmn}^{+,\times}e^{i\phi_{lmn}^{+,\times}}=\mathcal{A}_{lmn}e^{i\phi_{lmn}}\pm\mathcal{A}_{lmn}^{\prime}e^{-i\phi_{lmn}^{\prime}}$, where the signs $+(-)$ correspond to the $+(\times)$ polarizations respectively. The waveform $h$ detected by a detector is given by
\begin{align}
h&=h_+F_+(\theta_S,\phi_S,\psi_S,f)+h_\times F_\times(\theta_S,\phi_S,\psi_S,f),
\end{align}
where  $F_{+,\times}$ are frequency dependent pattern
functions (response functions) depending on the orientation
$\psi_S$ of the detector and the direction ($\theta_S,\phi_S$)
of the source. For LIGO (in the long wavelength limit), we have
\begin{subequations}
\begin{align}
F_{+}&=\frac{1}{2}(1+\cos^2\theta_{S})\cos2\phi_{s}\cos2\psi_{S}-\cos\theta_S\sin2\phi_S\sin2\psi_S,\\
F_{\times}&=\frac{1}{2}(1+\cos^2\theta_{S})\sin2\phi_{s}\cos2\psi_{S}+\cos\theta_S\sin2\phi_S\cos2\psi_S,
\end{align}
\end{subequations}
which are independent of  the frequency. The sky and polarization averaged
SNR is \cite{Cardoso,liuchang},
\begin{subequations}
\begin{align}
&\rho^2=4\int_0^{\infty}\frac{\langle\tilde{h}^{\ast}(f)\tilde{h}(f)\rangle}{S_N(f)}df=4\int _0^{\infty}\frac{|\tilde{h}_+|^2+|\tilde{h}_{\times}|^2}{S_n(f)}df\label{eq3},\\
&S_n(f)=\frac{S_N(f)}{\mathcal{R}(f)},\\
&\mathcal{R}(f)=\langle|F_+(f)|^2\rangle=\langle|F_{\times}(f)|^2\rangle,
\end{align}
\end{subequations}
where $\tilde{h}(f)$ is the Fourier  transform of the waveform,
$S_{N}(f)$ is the noise spectral density of the detector, $S_n(f)$
is the detector sensitivity, $\mathcal{R}(f)$ is the sky/polarization
averaged response function. The sky/polarization averaged  is defined by
\begin{equation}
\langle X \rangle=\frac{1}{4\pi^2}\int_0^{\pi}d\psi\int_0^{2\pi}d\phi\int_0^{\pi}X\sin\theta d\theta
\end{equation}
Especially, the response function $\mathcal{R}$
for LIGO is \cite{liuchang}
\begin{equation}
\mathcal{R}=\langle|F_+|^2\rangle=\langle|F_{\times}|^2\rangle=\frac{1}{32}\int_{-1}^{1}(1+6x^2+x^4)dx=\frac{1}{5},\quad x=-\cos\theta
\end{equation}
while the full  expressions of $F_+(f)$ and $F_{\times}(f)$ for LISA,
TianQin and TaiJi are much more complicated and can be found in \cite{Larson:1999we,Liang:2019pry}.
We perform the Fourier transform of the waveform by using the relation
\begin{equation}
\int_{-\infty}^{\infty}e^{i\omega t}\left(e^{\pm i\omega_{lmn}t-|t|/\tau_{lmn}}\right)dt=\frac{2/\tau_{lmn}}{(1/\tau_{lmn})^2+(\omega\pm\omega_{lmn})^2}\equiv 2b_{\pm}.\label{eq1}
\end{equation}
Based on Eq. (\ref{eq1}) we can easily work out  the Fourier transform of the plus and cross components,
\begin{subequations}\label{eq2}
\begin{align}
\tilde{h}_+&=\frac{1}{\sqrt{2}}\frac{M}{r}\mathcal{A}_{lmn}^{+}\left[e^{i\phi_{lmn}^{+}}S_{lmn}b_+ + e^{-i\phi_{lmn}^{+}}S_{lmn}^{\ast}b_-\right],\\
\tilde{h}_\times &=-\frac{1}{\sqrt{2}}\frac{iM}{r}\mathcal{A}_{lmn}^{\times}\left[e^{i\phi_{lmn}^{\times}}S_{lmn}b_+ - e^{-i\phi_{lmn}^{\times}}S_{lmn}^{\ast}b_-\right].
\end{align}
\end{subequations}
We add a correction factor $1/\sqrt{2}$ to serve as a compensation in amplitude because we are using the FH convention (developed by Flanagan and Hughes \cite{Flanagan:1997sx}) to calculate the SNR. In the FH convention, the waveform for $t<0$ is assumed to be identical to waveform for $t>0$ and therefore we can replace the decay factor $e^{-t/\tau_{lmn}}$ with $e^{-|t|/\tau_{lmn}}$ in the Fourier transform such that a compensation is needed for the doubling. Then we can insert Eq. (\ref{eq2}) into Eq. (\ref{eq3}) and do the integration to calculate SNR. However, as described in \cite{Cardoso}, a simple analytical formula of SNR can be derived by making  some approximations in  the calculation. In this way, we have the SNR expression as \cite{Cardoso},
\begin{equation}\label{eq4}
\rho_{FH}=\frac{1.31681\times10^4}{\mathcal{F}_{lmn}}\left(\frac{\epsilon_{rd}}{0.03}\right)^{\frac{1}{2}}\left(\frac{(1+z)M}{10^6M_{\odot}}\right)^{\frac{3}{2}}
\left(\frac{1\, \mathrm{Gpc}}{D_L(z)}\right)\left(\frac{S_0}{S_n(f_{lmn})}\right)^{\frac{1}{2}}\frac{2Q_{lmn}}{\sqrt{1+4Q_{lmn}^2}},
\end{equation}
where $S_0=1.59\times10^{-41}\mathrm{Hz}^{-1}$, $\mathcal{F}_{lmn}$ is the dimensionless frequency
defined by $\mathcal{F}_{lmn}=M\omega_{lmn}$, $\epsilon_{rd}$ is the radiation efficiency,
$M_{\odot}$ is the solar mass and $M$ is the black hole (source) mass,
$Q_{lmn}$ is a dimensionless quality factor of QNMs defined by
\begin{equation}
Q_{lmn}=\pi f_{lmn} \tau_{lmn}=\frac{1}{2}\omega_{lmn}\tau_{lmn},
\end{equation}
and $D_L(z)$ is the luminosity distance which can be  expressed as a function of cosmological redshift $z$ of the source in the standard flat $\Lambda$CDM cosmological model as
\begin{equation}
D_L(z)=\frac{1+z}{H_0}\int_0^{z}\frac{dz^{\prime}}{\sqrt{\Omega_{M}(1+z^{\prime})^3+\Omega_{\Lambda}}}.
\end{equation}
 We shall take the matter density $\Omega_M=0.32$, the dark energy
 density $\Omega_{\Lambda}=0.68$ and the Hubble constant $H_0=67\,\mathrm{km}\,\mathrm{s}^{-1}\mathrm{Mpc}^{-1}$.
 Eq. (\ref{eq4}) was derived in the context of LISA by making some approximations such as $S_{lmn}\simeq \mathfrak{R}(S_{lmn})$
 and large $Q_{lmn}$ limit. We will show that  these approximations and the derivation steps of Eq. (\ref{eq4})
 are not dependent on specific interferometric detectors, therefore Eq. (\ref{eq4}) could be applied to other space based detectors such as TianQin and TaiJi. We will discuss   the generality of the approximate SNR formula Eq. (\ref{eq4}) in more details in Appendix A,  and show that this formula is applicable to TianQin and TaiJi.

For the calculation of SNR, we will adopt the following noise and response functions for all three space based detectors \cite{Cornish:2001qi}
\begin{subequations}
\begin{align}
S_N(f)&=\frac{4S_a}{(2\pi f)^4 L^2}\left(1+\frac{10^{-4}\mathrm{Hz}}{f}\right)+\frac{S_x}{L^2},\\
\mathcal{R}(f)&=\frac{3}{10}\left[1+0.6\left(\frac{f}{f_{\ast}}\right)^2\right]^{-1},
\end{align}
\end{subequations}
in which $L$ is the detector arm length and $f_{\ast} = c/(2\pi L)$ is the transfer frequency, $S_a$ is the acceleration noise and $S_x$ is the position noise of the instruments, and  we list these  parameters for three detectors in Table. \ref{table0}.
\begin{table}[!htbp]
\centering
 \begin{tabular}{ccccccccccc}
    \hline\hline
    $ $ & LISA & TianQin & TaiJi\\
    \hline
    $L  $ & $2.5\times 10^9 \mathrm{m}$ & $\sqrt{3}\times 10^8 \mathrm{m}$ & $3\times10^9\mathrm{m}$ \\
    $\sqrt{S_a}$\, & $3\times10^{-15}\mathrm{ms}^{-2}/\mathrm{Hz}^{1/2}$\, & $10^{-15}\mathrm{ms}^{-2}/\mathrm{Hz}^{1/2}$\, & $3\times10^{-15}\mathrm{ms}^{-2}/\mathrm{Hz}^{1/2}$ \\
    $\sqrt{S_x}$ & $1.5\times10^{-11}\mathrm{m}/\mathrm{Hz}^{1/2}$ & $10^{-12}\mathrm{m}/\mathrm{Hz}^{1/2}$ & $8\times10^{-12}\mathrm{m}/\mathrm{Hz}^{1/2} $ \\
    \hline\hline
\end{tabular}
\caption{Parameters of all three space based detectors.\label{table0}}
\end{table}

In addition to the noise of the detectors, an effective  noise can be generated by the galactic binaries. For LISA,  the galactic noise can be well fitted as \cite{liuchang,Cornish2017}
\begin{equation}
S_c(f)=Af^{-7/3}e^{-f^{\alpha}+\beta f \sin(\kappa f)}[1+\tanh(\gamma(f_k-f))]\mathrm{Hz}^{-1},
\end{equation}
and the total sensitivity can be obtained by adding $S_c(f)$ to $S_n(f)$. The
effects of the galactic noise on the SNR for LISA  will be discussed  later,
and the parameters we are going to use for the four year mission lifetime
are $A=9\times 10^{-45}$, $\alpha=0.138$, $\beta=-221$, $\kappa=521$,
$\gamma=1680$, and $f_{k}=0.0013$ \cite{liuchang}.

\begin{figure}[htp]
\centering
\includegraphics[height=3in,width=4.8in]{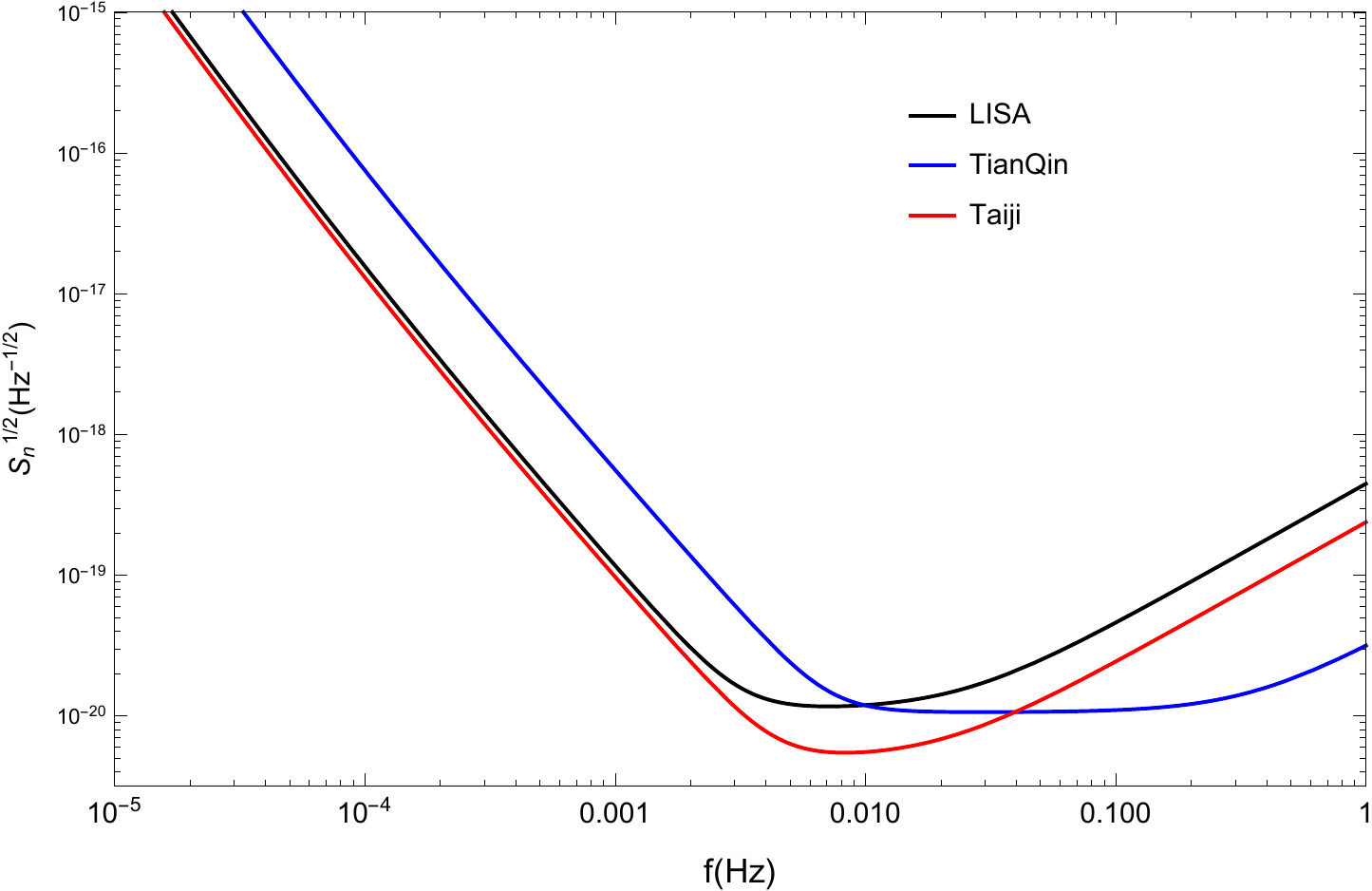}
\caption{The root sensitivity curves for LISA, TianQin and TaiJi.}
\label{fig1}
\end{figure}

We show the root sensitivity curve for LISA, TianQin and TaiJi in Fig. \ref{fig1},  from which we can see that the sensitivity value of TianQin is higher than that of LISA in the frequency range $f\lesssim0.01\mathrm{Hz}$ and the sensitivity value of TianQin can be higher than Taiji when $f\lesssim0.04\mathrm{Hz}$. For the rest regions of frequency respectively, the LISA and TaiJi have higher sensitivity than that of TianQin, which suggests that comparing to TianQin, LISA and TaiJi are better for gravitational wave detections at lower frequencies (usually corresponds to higher black hole mass), while for higher frequencies we should turn to count on TianQin. In addition, we can see that the sensitivity of TaiJi is always lower than that of LISA in the whole frequency band which implies that TaiJi can be more sensitive to detect the gravitational wave signals emitted from the same source when compared with LISA.

\section{SNR for non-singular BHs in conformal gravity}
\label{section3}

\subsection{Quasinormal modes of  non-singular BHs in conformal gravity}

The metric of non-singular  BHs in conformal gravity can be expressed as \cite{Bambi1}
\begin{equation}
ds^2=S(r)ds_{\mathrm{Schw}}^2=-S(r)f(r)dt^2+\frac{S(r)}{f(r)}dr^2+S(r)r^2d\Omega^2, \label{eq6}
\end{equation}
where $ds_{\mathrm{Schw}}^2$ is the Schwarzschild spacetime line element,
with $f(r)=1-2M/r$, the factor $S(r)$  is  \cite{Bambi1}
\begin{equation}
S(r)=\left(1+\frac{L^2}{r^2}\right)^{2N},
\end{equation}
and $N$ is an arbitrary positive integer and $L$ is a new length scale. The additional conformal
factor $S(r)$ making the spacetime singularity-free distinguishes the metric (\ref{eq6})
from the Schwarzschild metric, and  the metric \eqref{eq6} can reduce to Schwarzschild form when $S(r)\rightarrow 1$,
i.e. $N\rightarrow 0$ or $L\rightarrow 0$. We will show that the non-zero parameters $N, L$ will influence the dynamical behavior reflected by QNMs of  BHs under gravitational perturbations. Since QNMs can disclose the black hole fingerprint, it can differentiate such non-singular  BHs from the Schwarzschild black hole, as we will discuss in the following.

The master equation for the axial gravitational perturbation reads \cite{Chen:2019iuo}
\begin{equation}
\frac{d^2H^{(-)}}{dr_{\ast}^2}+(\omega^2-V(r))H^{(-)}=0,
\end{equation}
where we have used tortoise radius defined by $dr/dr_{\ast}=f(r)$,
 $H^{(-)}$ is the radial part of the axial gravitational perturbation,
$\omega$ is the QNM frequency, the effective potential $V$ is \cite{Chen:2019iuo}
\begin{equation}\label{eq10}
V(r)=f(r)\left\{\frac{l(l+1)}{r^2}-\frac{2}{r^2}-Z\frac{d}{dr}\left(\frac{f(r) dZ/dr}{Z^2}\right)\right\},
\end{equation}
and $Z(r)=\sqrt{S(r)}r$.

In \cite{Chen:2019iuo}, the 6th order WKB method was adopted to compute the QNM of the non-singular black hole configuration. In our numerical computation, we employ the 13th order WKB approximation. In the study of gravitational perturbations in the Schwarzschild black hole \cite{Matyjasek:2017psv}, comparing with the accurate numerical result, it was found that the 13th order WKB is more precise than the 6th WKB approach. It was argued that the WKB approximation works in satisfactory accuracy in calculating the QNM once $l>> n$ \cite{Iyer:1986np}, while does not work well for high overtone modes. In \cite{Chen:2019iuo} the discussion on the QNM was only limited to the lowest QNM for the gravitational perturbation. Including the Pade approximation, it was observed that there is a great increase of accuracy in calculating the QNM by using the WKB approach, furthermore with averaging of the Pade approximation accurate calculations can be achieved not only in the lowest mode, but also for the overtone modes with  slightly bigger $n$ than $l$, however the numerical results are still not much reliable for $n\gg l$,  even if the Pade approximation is included \cite{Matyjasek:2017psv}.

In our numerical computation, we employ the 13th order WKB approximation method with averaging of the Pade approximations \cite{Matyjasek:2017psv} to
calculate the QNMs of the axial gravitational perturbation on the background of
non-singular BHs. We show our results in  Tables \ref{table1}
and  \ref{table2} where we express the QNM frequency in a
dimensionless variable $M\omega$.


\begin{table}[!htbp]
\centering
\caption{Quasinormal frequency $M\omega$ for non-singular  BHs in conformal gravity for parameter $L=2$.}
        \begin{tabular}{ccccccccccc}
    \hline\hline
    $l$ & $n$ & $N=0$  & $N=2$ & $N=5$ & $N=10$ \\
    \hline
    $2$&$0\,$  &$0.373675-0.088964i\,$  &$0.409160-0.121392i\, $&$0.715683-0.134492i\, $&$1.372383-0.139325i $\\
    $ $&$1$&$0.346827-0.273930i $&$0.407758-0.370858i $&$0.719270-0.403565i $&$1.376117-0.417754i $\\
    $ $&$2$&$0.299998-0.478098i $&$0.414498-0.620991i $&$0.727050-0.672962i $&$1.383599-0.695444i $\\
    $3$&$0$&$0.599443-0.092703i $&$0.620553-0.107015i $&$0.855703-0.125041i $&$1.450989-0.135461i $\\
    $ $&$1$&$0.582643-0.281297i $&$0.604412-0.331714i $&$0.853220-0.377014i $&$1.452762-0.406564i $\\
    $ $&$2$&$0.551686-0.479087i $&$0.589888-0.569715i $&$0.851142-0.632531i $&$1.456717-0.678002i $\\
    $ $&$3$&$0.511943-0.690318i $&$0.584544-0.815114i $&$0.851857-0.890962i $&$1.463341-0.949499i $\\
    $4$&$0$&$0.809178-0.094163i $&$0.825032-0.102217i $&$1.013367-0.117947i $&$1.549660-0.131294i $\\
    $ $&$1$&$0.796631-0.284334i $&$0.810029-0.311229i $&$1.007860-0.356078i $&$1.549669-0.394338i $\\
    $ $&$2$&$0.772709-0.479908i $&$0.786423-0.534230i $&$0.999971-0.599176i $&$1.550292-0.658511i $\\
    $ $&$3$&$0.739836-0.683924i $&$0.766779-0.768643i $&$0.992948-0.846952i $&$1.552405-0.923937i $\\
    $ $&$4$&$0.701514-0.898237i $&$0.754662-1.007982i $&$0.988570-1.097952i $&$1.556748-1.190222i $\\
    $5$&$0$&$1.012295-0.094870i $&$1.025096-0.100075i $&$1.181610-0.112849i $&$1.664843-0.127206i $\\
    $ $&$1$&$1.002221-0.285817i $&$1.013275-0.302492i $&$1.174978-0.340491i $&$1.663474-0.382215i $\\
    $ $&$2$&$0.982695-0.480328i $&$0.991012-0.512278i $&$1.164029-0.573071i $&$1.661378-0.638801i $\\
    $ $&$3$&$0.955004-0.680556i $&$0.963850-0.734449i $&$1.152084-0.811398i $&$1.659565-0.897424i $\\
    $ $&$4$&$0.921081-0.888197i $&$0.940140-0.967011i $&$1.141652-1.054348i $&$1.659055-1.157978i $\\
    $ $&$5$&$0.883335-1.104182i $&$0.840158-1.120908i $&$1.133807-1.300278i $&$1.660604-1.419940i $\\
    \hline\hline
    \label{table1}
\end{tabular}
\end{table}

\begin{table}[!htbp]
\centering
\caption{Quasinormal frequency $M\omega$ for non-singular  BHs in conformal gravity for parameter $N=2$.}
        \begin{tabular}{ccccccccccc}
    \hline\hline
    $l$ & $n$ & $L=0$& $L=2$ & $L=5$ & $L=15$ \\
    \hline
    $2$&$0\,$&$0.373675-0.088964i\, $&$0.409161-0.121392i\, $
    &$0.500121-0.105834i\, $&$0.530178-0.094147i $\\
    $ $&$1$&$0.346827-0.273931i $&$0.407758-0.370858i $&$0.496880-0.317166i $&$0.514476-0.284905i $\\
    $ $&$2$&$0.299998-0.478098i $&$0.414498-0.620991i $&$0.489883-0.526622i $&$0.484316-0.483647i $\\
    $3$&$0$&$0.599443-0.092703i $&$0.620553-0.107015i $&$0.686818-0.101541i $&$0.708826-0.094852i $\\
    $ $&$1$&$0.582643-0.281297i $&$0.604412-0.331714i $&$0.679594-0.305861i $&$0.696048-0.286379i $\\
    $ $&$2$&$0.551686-0.479087i $&$0.589888-0.569715i $&$0.667081-0.512816i $&$0.671468-0.483427i $\\
    $ $&$3$&$0.511943-0.690318i $&$0.584544-0.815114i $&$0.648889-0.720679i $&$0.637497-0.689529i $\\
    $4$&$0$&$0.809178-0.094163i $&$0.825032-0.102217i $&$0.876184-0.099544i $&$0.893459-0.095293i $\\
    $ $&$1$&$0.796631-0.284334i $&$0.810029-0.311229i $&$0.868380-0.299798i $&$0.882860-0.287183i $\\
    $ $&$2$&$0.772709-0.479908i $&$0.786423-0.534230i $&$0.853993-0.502905i $&$0.862299-0.482970i $\\
    $ $&$3$&$0.739836-0.683924i $&$0.766779-0.768643i $&$0.834321-0.709301i $&$0.833181-0.685113i $\\
    $ $&$4$&$0.701514-0.898237i $&$0.754662-1.007982i $&$0.809681-0.918896i $&$0.797793-0.895593i $\\
    $5$&$0$&$1.012295-0.094871i $&$1.025096-0.100075i $&$1.066719-0.098481i $&$1.080916-0.095562i $\\
    $ $&$1$&$1.002221-0.285817i $&$1.013275-0.302492i $&$1.059305-0.296361i $&$1.071930-0.287641i $\\
    $ $&$2$&$0.982695-0.480328i $&$0.991012-0.512278i $&$1.045228-0.496726i $&$1.054373-0.482565i $\\
    $ $&$3$&$0.955004-0.680556i $&$0.963851-0.734449i $&$1.025583-0.700485i $&$1.029119-0.682139i $\\
    $ $&$4$&$0.921081-0.888197i $&$0.940140-0.967011i $&$1.001235-0.907965i $&$0.997555-0.887945i $\\
    $ $&$5$&$0.883335-1.104182i $&$0.840158-1.120908i $&$0.972539-1.119446i $&$0.961550-1.101097i $\\
    \hline\hline
    \label{table2}
\end{tabular}
\end{table}

In Table. \ref{table1} where we fix the parameter $L$, we find that with the increase of $N$  both the real part representing the oscillation frequency and the magnitude of the imaginary part relating to the damping time of QNMs will increase, which implies that with the increase of the parameter $N$ in the non-singular black hole in conformal gravity, the gravitational perturbation can have more oscillations but die out faster. Comparing to the non-singular black hole backgrounds, we find that the perturbation of the Schwarzschild black hole with $N=0$ can last longer. Our result confirms that reported in \cite{Chen:2019iuo} where they limited their discussion to a fixed angular index $l$. Since we have adopted the Pade approximation, we can accurately calculate QNMs for the change of $n, l$ until $n=l$ (to keep numerical accuracy, $n>l$ is not considered in our discussion). In the Schwarzschild background, for the same overtone mode  when the angular index $l$ becomes higher, we observe that the real parts of frequency are always higher, while the imaginary part is higher for $n\leq2$, but  decreases when $n>2$. However this property does not hold for non-singular  BHs with $N\neq0$.  In non-singular holes, for the same overtone modes the higher angular number $l$ always results in a higher real part of the frequency but smaller imaginary part of the frequency,  which suggests that for the same overtone mode the perturbation with higher angular index may last longer for non-singular  BHs while in
the Schwarzschild black hole perturbation the mode $l=2,n=0$ is always the  longest one.

In Table. \ref{table2} we present the frequencies of QNMs for a fixed $N$ parameter.
With the increase of $L$, the real part of QNMs  monotonously increases while the imaginary part increases
from $L=0$ to $L=2$ but then decreases continuously with the further increase of $L$. This behavior agrees to the result reported in \cite{Chen:2019iuo} for a fixed angular index $l$. Employing the Pade approximation,
we accurately calculated QNMs with our 13th WKB approach for different $n, l$ even when $n=l$.
Similar to the Schwarzschild black hole, for the non-singular BHs we find that for  the same overtone modes,
with the increase of the angular number $l$, the real part of the frequency increases.
The absolute imaginary part of the frequency for non-singular  BHs presents different behaviors
from that of the Schwarzschild background when $L$ is not big enough.
For the same overtone mode, with the increase of the angular number $l$,
the absolute imaginary part of the frequency for a non-singular  black hole decreases
instead of  increasing as in the Schwarzschild background.
This is consistent with the picture we learn from Table.\ref{table1},
for the same overtone mode the perturbation for a non-singular black hole with
a larger angular index $l$ may last longer, which is different from the case in the Schwarzschild background,
where the fundamental mode $n=0,l=2$ always dominates.
The result of changing $L$ looks more complicated than that for the change of $N$ given above.
When $L=15$, the QNM frequencies return to the similar behavior with the change of $n, l$ to that in the Schwarzschild background.
In this case, for the same overtone number,
we can see that the real part of the frequency increases monotonously with the angular number $l$,
while for $n\leq1$ the imaginary part is higher for larger $l$, but for $n>1$ it decreases when increasing $l$.

Precise numerical results of the QNM frequencies for different $(n,l)$ are useful to calculate the multi-mode SNR of this non-singular black hole. However in this work we will concentrate on the single-mode SNR.
Different from the Schwarzschild black hole, it looks that in the non-singular black hole background
the mode $n=0, l=2$ is not apparently the dominant mode (here `dominant mode' means the mode with slowest damping rate corresponding to smallest value of $-\mathrm{Im}\, \omega$).
Instead,  for the same overtone mode, the imaginary frequency for a bigger angular index implies that the perturbation may last longer in the non-singular black hole.
However, we notice that in the limit $l\rightarrow\infty$, the effective potential $V(r)\approx f(r){l^2}/{r^2} $,
which reduces to that of Schwarzschild  BHs,
which makes it difficult to distinguish the modes between the non-singular black hole and the
Schwarzschild black hole in the large $l$ limit. In the face of complicated data,
actually a criterion to determine the dominant mode was suggested in \cite{Wang:2004bv}. We redefine the dominant mode by choosing min$\{\sqrt{\omega_R^2+\omega_I^2}\}$. Applying this criterion, we find that it is always the $n=0,l=2$ mode that serves the dominant mode in the perturbation, which holds also in the non-singular black hole. If we look at the relation Eq. (\ref{eq5}) between the GW amplitude $\mathcal{A}_{lm}$ and the energy radiation efficiency $\epsilon_{rd}$,  the $n=0,l=2$ mode always has the strongest amplitude corresponding to more powerful energy in this mode. This further guarantees that the $n=0,l=2$ mode dominates in the perturbation in the non-singular black hole.  Now we can compare the same dominant single-mode SNR for the non-singular black hole and the Schwarzschild black hole,
which allows us to explore their imprints in GWs.

\subsection{SNR by LISA, TianQin and TaiJi}

In this subsection we calculate the SNR for LISA, TianQin and TaiJi by using the QNMs we have obtained  for non-singular  BHs in conformal gravity. We explore the SNR related to the dominant mode $n=0,l=2$ in both of the non-singular and Schwarzschild  BHs.  At first we would like to focus on the discussion of SNR for LISA, and then we will take TianQin and TaiJi into consideration for comparisons. In the calculation of SNR, we would like to set an optimistic value of radiation efficiency $\epsilon_{rd}=3\%$ assumed in \cite{Flanagan:1997sx}, which is based on  quadrupole-formula-based estimate of the QNMs amplitude when the distortion of the horizon of the black hole is of order unity \cite{Flanagan:1997sx}, as well as  a pessimistic  value $\epsilon_{rd}=0.1\%$ corresponding to the estimates for the energy emitted  in the head-on collision of equal-mass black holes \cite{Sperhake:2005uf}.

\begin{figure}[thbp]
\centering
\includegraphics[height=2.2in,width=3.2in]{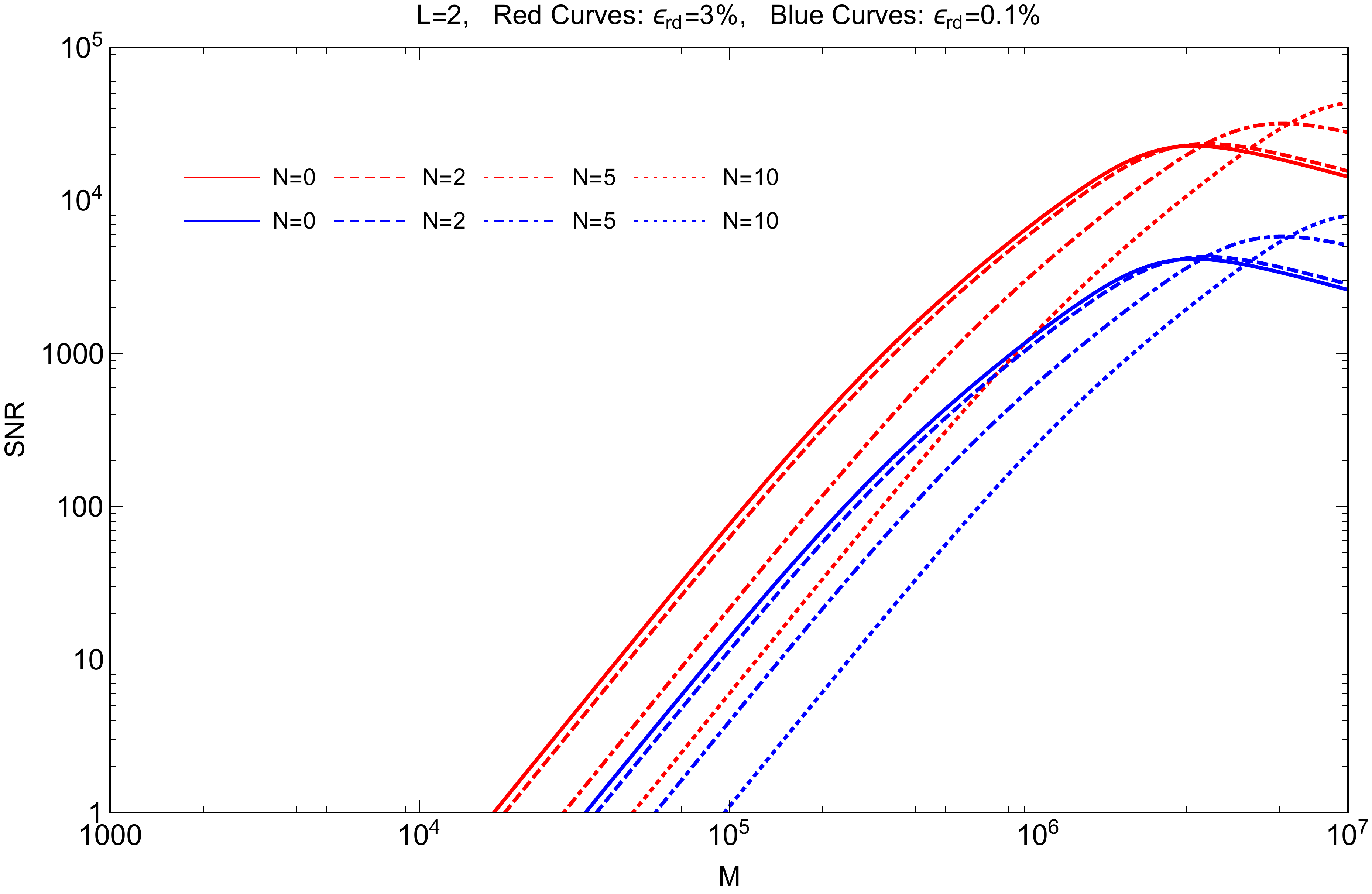}
\includegraphics[height=2.2in,width=3.2in]{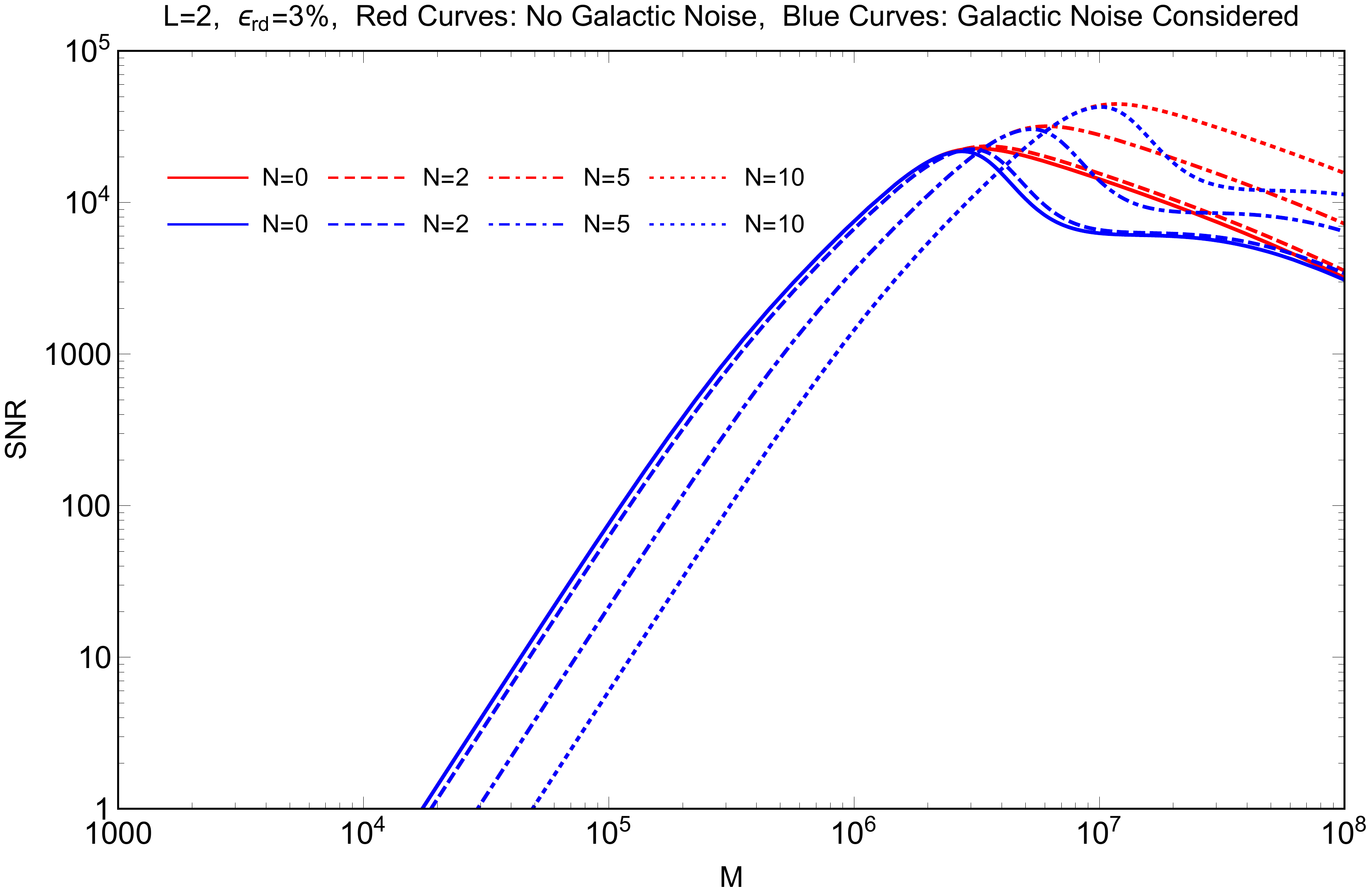}
\caption{The SNR behavior of LISA with the change of the black hole mass $M$ at a distance $D_L=3\,\mathrm{Gpc},z=0.54$ with the angular number $l=2$ in QNMs. For each plot we have fixed the value of $L$ but changing $N$. For the left plot,  the red curves and blue curves correspond to the radiation efficiency $\epsilon_{rd}=3\%$ and $0.1\%$, respectively, while for the right plot  we set $\epsilon_{rd}=3\%$, and  the red curves denote the SNR without including the galactic noise and the blue curves denote the SNR affected by the galactic noise.\label{fig2}}
\end{figure}

\begin{figure}[thbp]
\centering
\includegraphics[height=2.2in,width=3.2in]{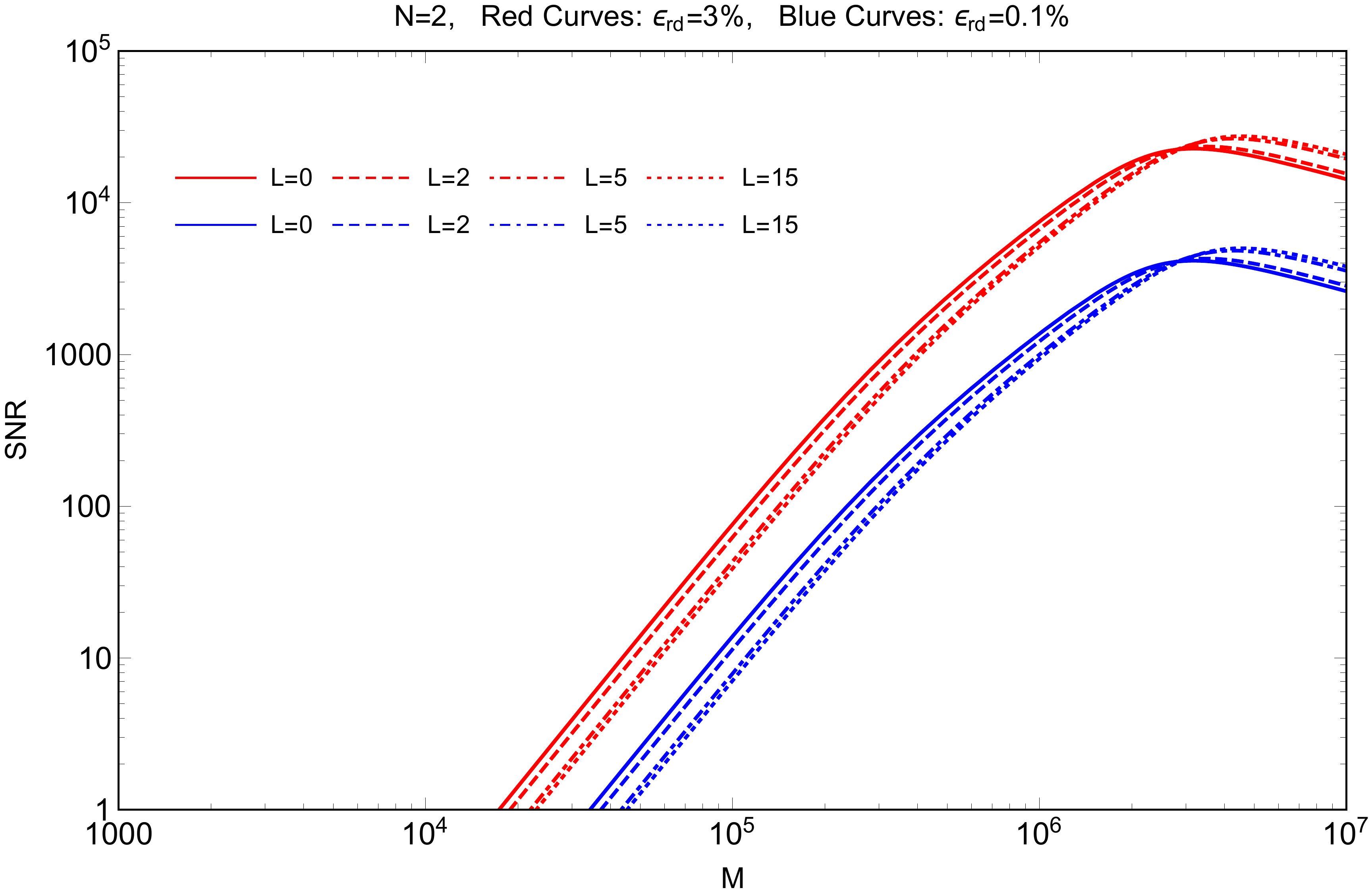}
\includegraphics[height=2.2in,width=3.2in]{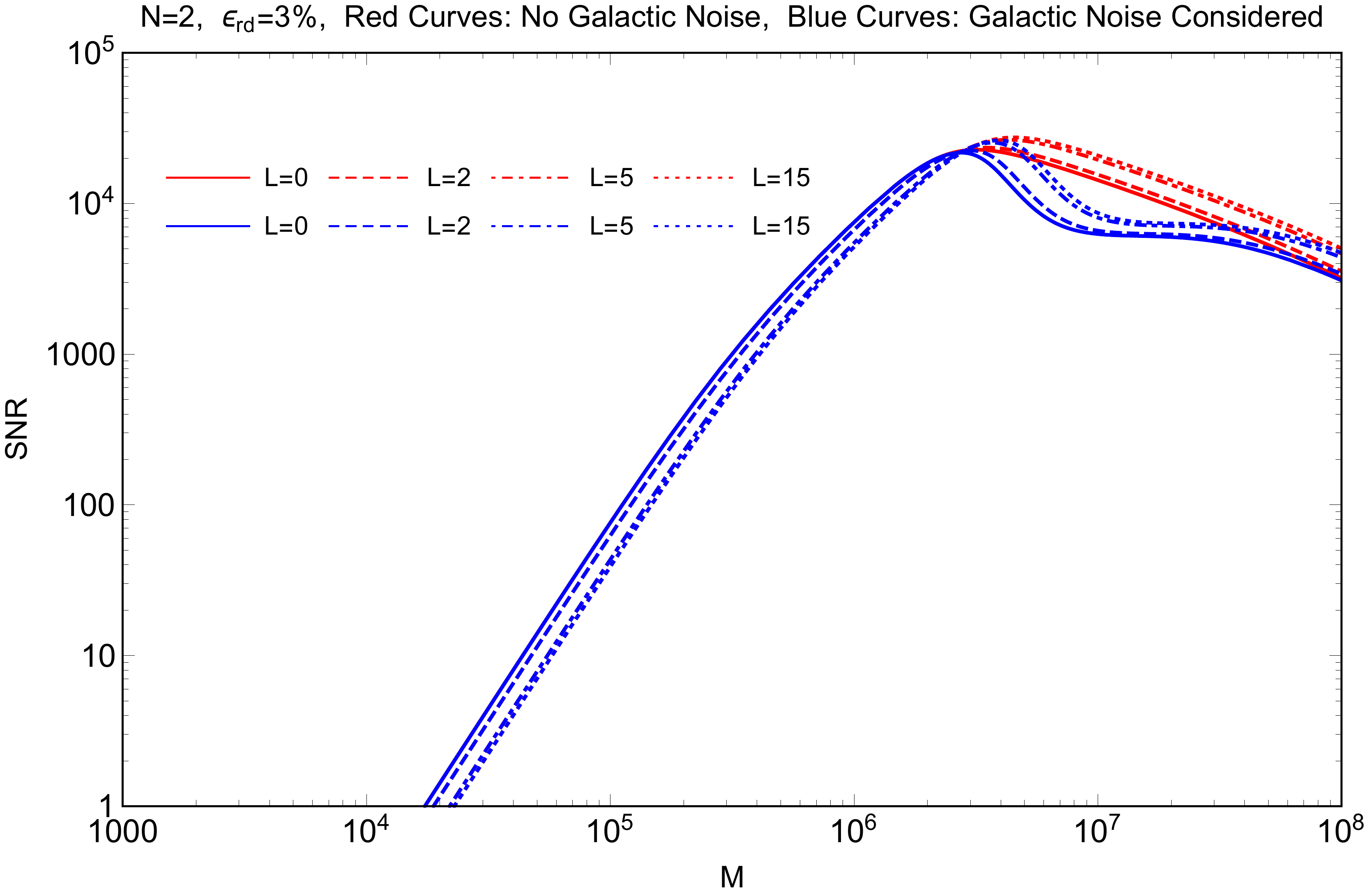}
\caption{The SNR behavior of LISA with the change of the black hole mass $M$ at a distance $D_L=3\,\mathrm{Gpc},z=0.54$  with the angular number $l=2$. For each plot we have fixed the value of $N$ but change $L$. For the  plot on the left, the galactic noise is not included and the red curves and blue curves correspond to the radiation efficiency $\epsilon_{rd}=3\%$ and $0.1\%$, respectively, while for the  plot on the right we set $\epsilon_{rd}=3\%$, and  the red curves denote the SNR without including the galactic noise and the blue curves denote the SNR affected by the galactic noise.\label{fig3}}
\end{figure}

In Fig. \ref{fig2} we show the SNR curve  by fixing the parameter $L$, while changing the parameter $N$.
We can see that with the increase of the mass, the SNR will grow, which implies that LISA is more sensitive to GW signals generated by  BHs with greater mass. For the Schwarzschild black hole with $N=0$,
the SNR will reach the maximum when the black hole mass becomes $\sim10^6M_{\odot}$.
Thus for the Schwarzschild black hole LISA is most sensitive when the black hole mass is around $2\times 10^6 M_{\odot}$.
Considering the non-singular black hole with bigger $N$,
we see that the SNR is smaller than that of the Schwarzschild black hole when the black hole mass is below $2\times 10^6 M_{\odot}$ and with the increase of $N$, the SNR is more suppressed when the black hole mass is within this value.
However when the black hole is more massive, the SNR of non-singular  BHs catches up and exceeds
further the value of the Schwarzschild black hole.
In Fig. \ref{fig3} we show the SNR at a fixed parameter $N$ but with  changing of the parameter $L$ in each plot.
The general feature in this case is similar to that illustrated in Fig. \ref{fig2},
the non-singular black hole has higher SNR when the black hole becomes more massive.

The radiation efficiency plays an important role in the SNR. As a natural result, one can find that higher radiation efficiency leads to higher SNR since $\rho\sim\sqrt{\epsilon_{rd}}$ and this fact is reflected by Eq. (\ref{eq4}). As above disclosed, within a certain mass region, the SNR of non-singular black holes is decreased when increasing parameter $L$ and $N$. Concerning this  effect, a question may arise about the detectability of non-singular black holes, and it is  assumed SNR $\rho\geq10$ as a criterion for detectability  as suggested in \cite{Cardoso}. Follow this criterion, it is encouraging to see that even in the situation of a pessimistic head-on collision with $\epsilon_{rd}=0.1\%$, we can still have SNR $\rho\geq10$ for non-singular black holes mass $M\gtrsim 10^5M_{\odot}$, which means that non-singular black holes are reasonably  expected to be detected by LISA.

When the galactic confusion noise is taken into consideration in Fig. \ref{fig2} and Fig. \ref{fig3},
we see a dip   appears when the black hole mass is a few  times of $10^6M_{\odot}$, for non-singular  BHs with bigger $L$ and $N$ the dip starts to  appear for more massive  BHs. For smaller masses, the effect on SNR is negligible. This result indicates that the non-singular black hole model allows a wider mass range in which the SNR can avoid the influence of galactic noise. As a possible consequence, in the detection of the ringdown signals of the  super massive black holes centered at our galaxy with the mass estimated around $M\simeq 3.7\pm0.2\times10^6M_{\odot}$, the SNR may not be lowered by the galactic noise if that black hole is non-singular, whereas in the case of singular black holes predicted by general relativity, the SNR will probably be impacted.

\begin{figure}[thbp]
\centering
\includegraphics[height=2.2in,width=3.2in]{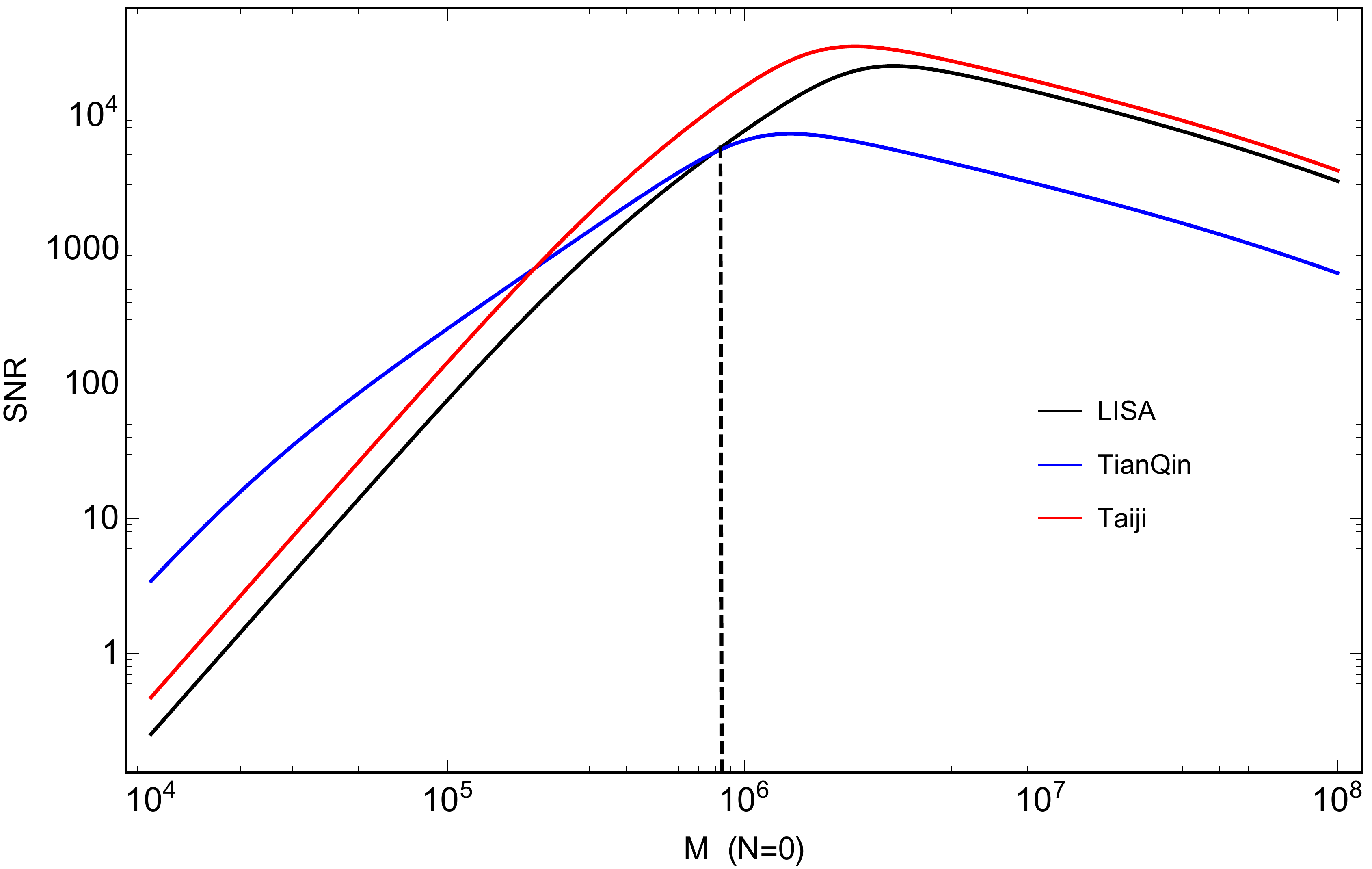}
\includegraphics[height=2.2in,width=3.2in]{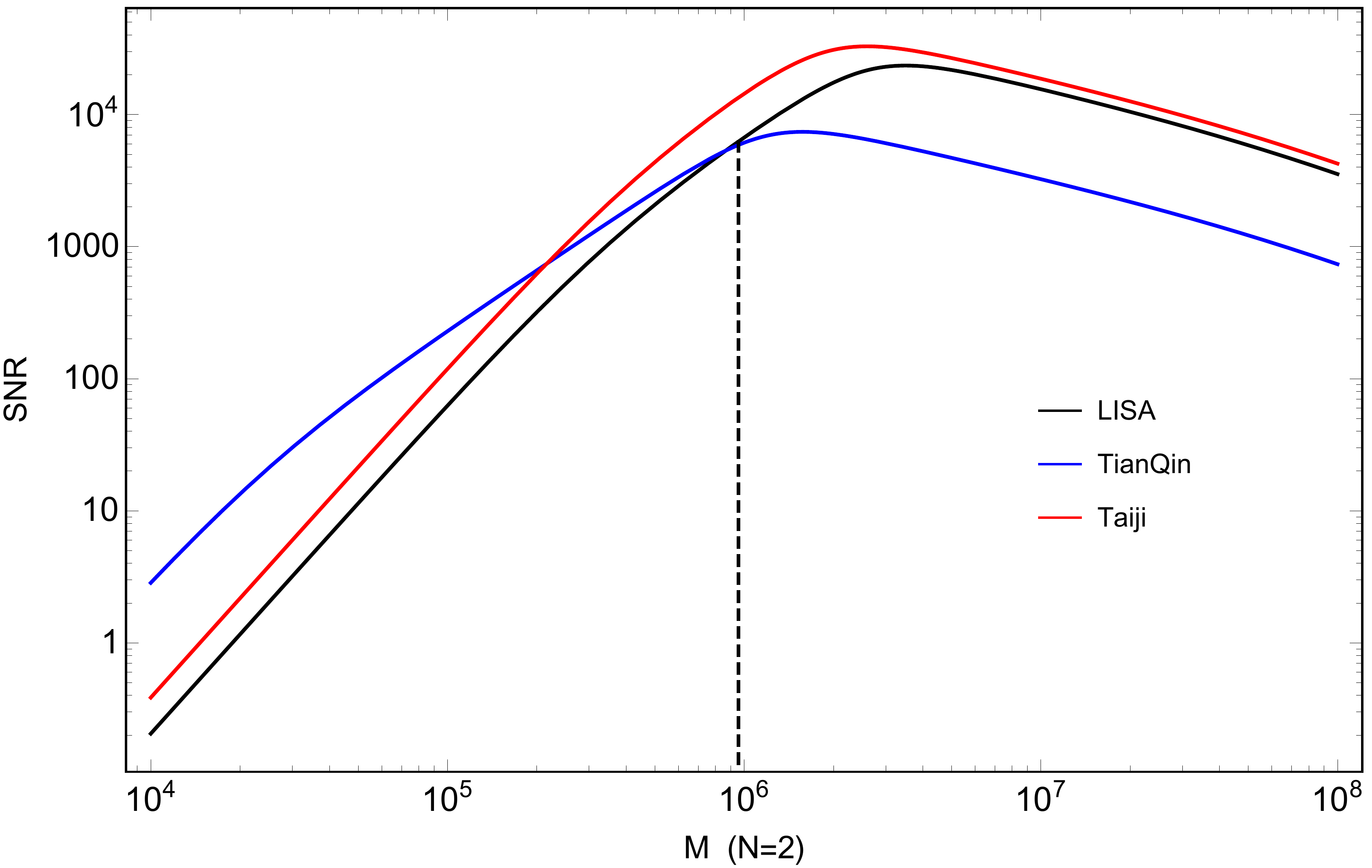}
\includegraphics[height=2.2in,width=3.2in]{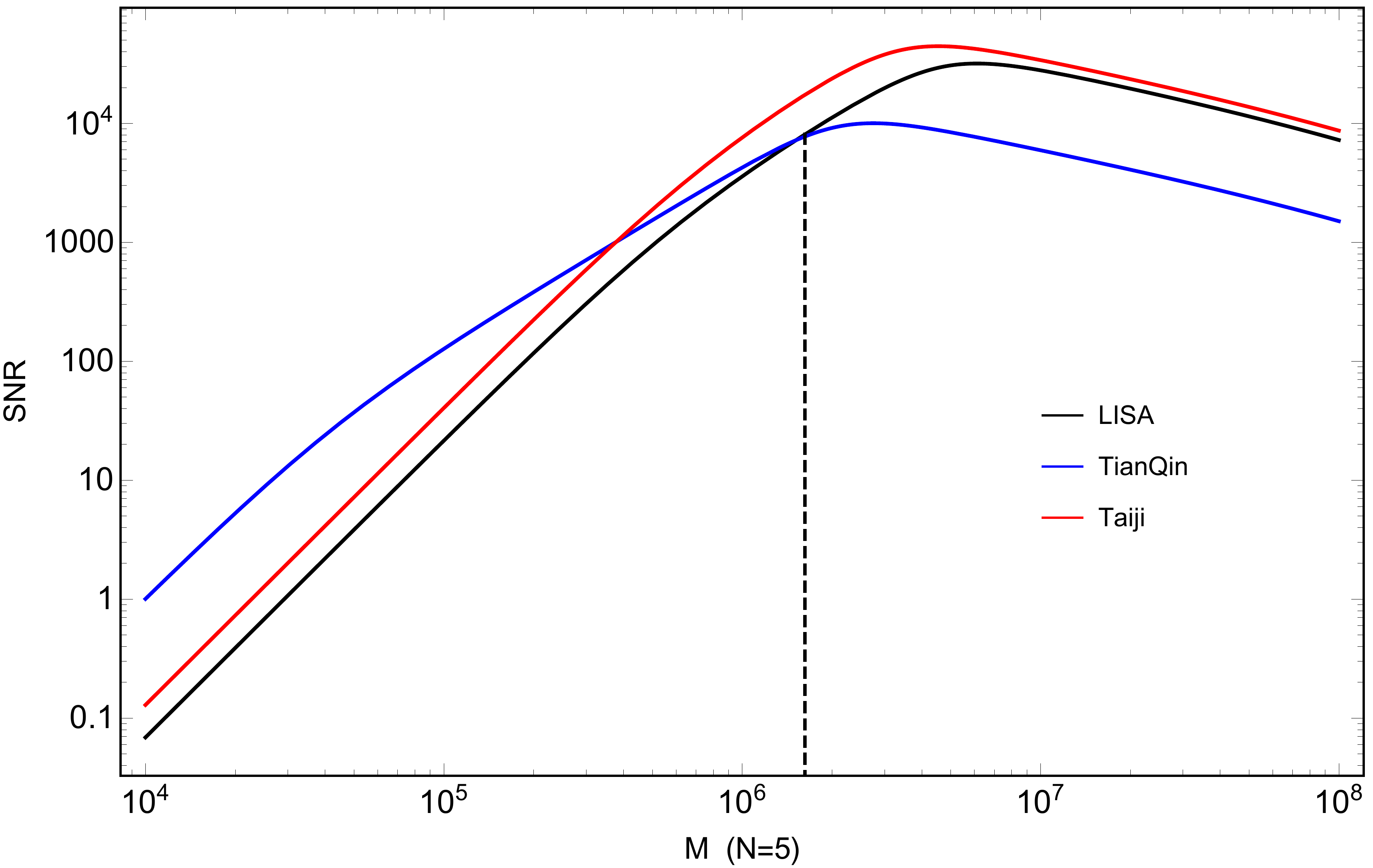}
\includegraphics[height=2.2in,width=3.2in]{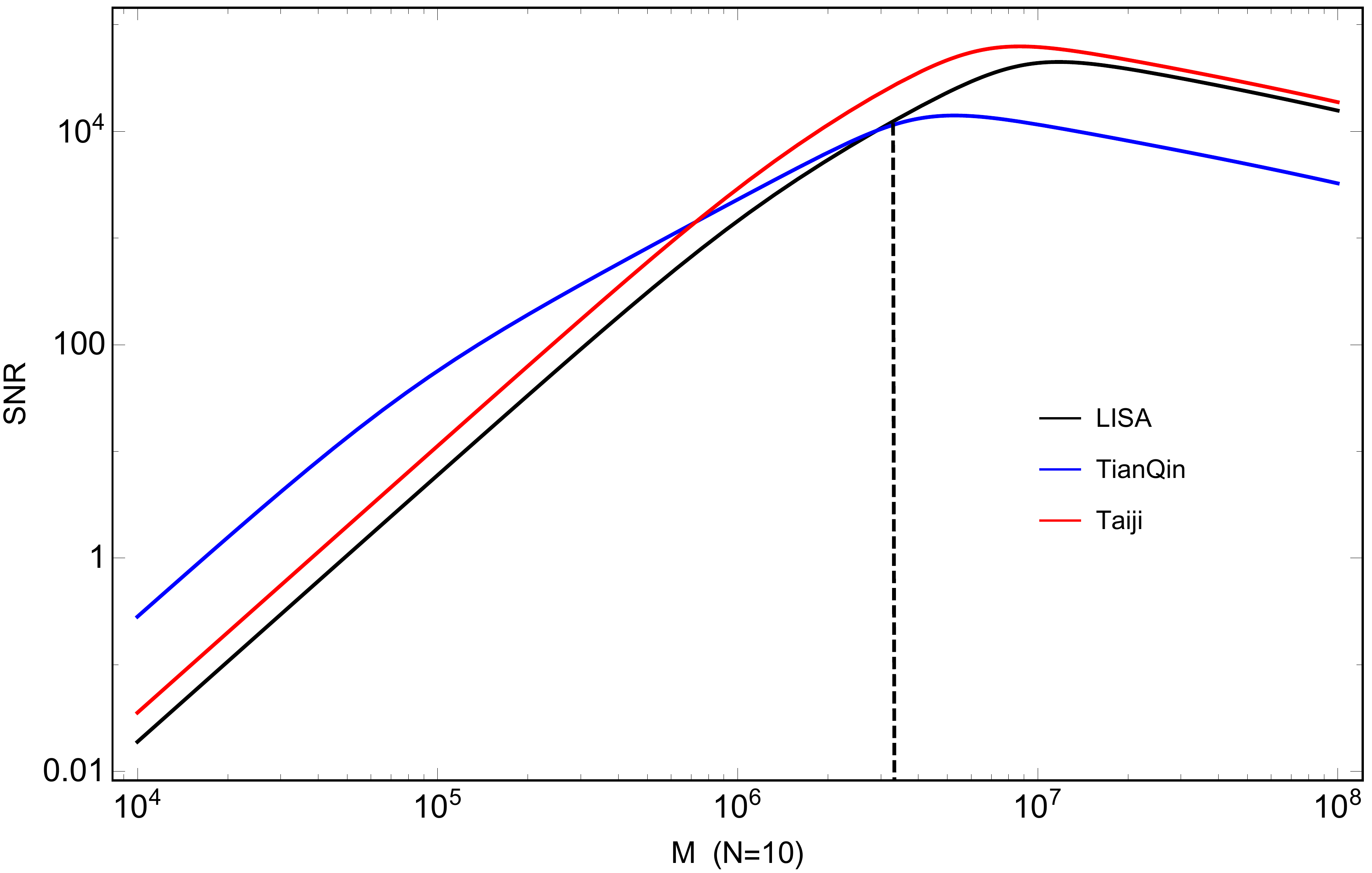}
\caption{The comparison of SNR  among LISA, TianQin and TaiJi with different  black hole masses  at a distance $D_L=3\,\mathrm{Gpc},z=0.54$ by taking the angular number $l=2$ and the radiation efficiency $\epsilon_{rd}=0.03$. In each plot we set $L=2$, and four different $N$ are designated correspondingly in the four plots.\label{fig4}}
\end{figure}

In Fig. \ref{fig4} we show a comparison of SNR among LISA, TianQin and TaiJi.
From this figure one can see that there exists a maximal value of SNR for all these three detectors,
and the mass related to the maximal SNR grows when the  BHs deviate extensively from the Schwarzschild ones.
It is clear to see  that there exists a critical mass $M_{cri}$ in each plot.
For the mass range $M<M_{cri}$, the SNR for TianQin is higher than LISA implying that TianQin is more sensitive to GWs emitted from  BHs with comparatively smaller masses,
while for more massive  BHs with $M>M_{cri}$ the LISA and TaiJi is more sensitive for the detection.
Different sensitivities of these three detectors were also reflected in the root sensitive curve shown in Fig. \ref{fig1} which demonstrated that LISA and TaiJi are more sensitive to lower frequency (corresponding to bigger  BHs) and  TianQin is more sensitive to comparatively higher frequency GW signals (corresponding to smaller  BHs). It is interesting to note that the critical mass $M_{cri}$ is related to the parameter $N$,
which increases when the black hole deviates more from the standard Schwarzschild black hole.
It is noticeable that in the whole frequency band (from low frequency to high frequency) the SNR of TaiJi is higher than LISA,
which is consistent with the sensitivity demonstrated in Fig. \ref{fig1}.
Comparing the values of $M_{cri}$ and the locations of the dominant mode peaks of SNR for different non-singular parameters,
LISA and TaiJi are more  promising to distinguish non-singular  BHs from the standard Schwarzschild ones.

\section{SNR for non-singular Bardeen BHs}
\label{section4}

\subsection{Quasinormal modes of Bardeen BHs}

The metric of the non-singular Bardeen black hole is \cite{Bardeen}
\begin{equation}
ds^2=-f(r)dt^2+\frac{1}{f(r)}dr^2+r^2(d\theta^2+\sin\theta^2d\phi^2),
\end{equation}
where $f(r)$ is given by \cite{Bardeen}
\begin{equation}
f(r)=1-\frac{2Mr^2}{(r^2+\beta^2)^{\frac{3}{2}}}.
\end{equation}
The parameter $\beta$ can be regarded as the charge of a self-gravitating magnetic monopole system with mass $M$.
To ensure the existence of  BHs, the parameter $\beta$ must be restricted to be $\beta^2\leq \frac{16}{27}M^2$ and one can clearly see that when $\beta=0$ the metric reduces to the Schwarzschild black hole.
This parameter $\beta$ makes the spacetime non-singular, which leads to  different dynamical behaviors of the gravitational perturbation in contrast to that of the  Schwarzschild black hole.

The master equation for the axial gravitational perturbation was given by \cite{Ulhoa}
\begin{equation}
\frac{d^2\phi}{dr_{\ast}^2}+\left[\omega^2-V(r)\right]\phi=0,
\end{equation}
where the effective potential $V(r)$ reads
\begin{equation}
V(r)=f(r)\left(\frac{l(l+1)+2(f(r)-1)}{r^2}+\frac{1}{r}\frac{df(r)}{dr}+\frac{d^2f(r)}{dr^2}+2\kappa L\right),
\end{equation}
in which $\kappa=8\pi$, and
\begin{equation}
L=\frac{3M}{|\beta|^3}\left(\frac{\sqrt{2\beta^2F}}{1+\sqrt{2\beta^2F}}\right)^{\frac{5}{2}},\qquad F=\frac{\beta^2}{2r^4}.
\end{equation}

In \cite{Ulhoa} the QNM was calculated by using the  3rd WKB method.
It was found that compared with high order WKB approaches,
the numerical result obtained by the  3rd WKB method is
not very accurate \cite{Konoplya:2003ii}.
In order to distinguish this non-singular black hole from the Schwarzschild black hole, we need very accurate results of the QNM spectrum. Therefore,  in our calculations we will employ the 13th order WKB method and the Pade approximation to guarantee the high precision in our numerical computation.

We list our result in Table. \ref{table3}. Analyzing the frequency of QNMs,
we learn that with the increase of $\beta$, the real part of the QNM frequency increases for every fixed $l, n$ mode,
while the imaginary part of the perturbation frequency decreases for any given  $l>2$ with different $n$.
Our result is different from that in \cite{Ulhoa}, where it was claimed that the imaginary frequency keeps almost the same for different choices of $\beta$.
This is because their  3rd WKB method is not accurate enough  to show the details.
Moreover in \cite{Ulhoa} the behavior of QNMs with the change of the angular number $l$ is not discussed.
With the Pade approximation, we are in a position to analyze carefully the dependence of the QNM frequency on the angular index $l$ and the overtone number $n$ until the limit $n\sim l$.
 With the increase of $l$ at the same overtone number $n$,
we find that both the real part and the imaginary part monotonously increase for $\beta=0$ and $\beta=0.3$ in the condition $n\leq2$ and $n\leq1$, respectively.
For bigger $\beta$, for example $\beta=0.6$, the imaginary part behaves differently.
We have the spectrum of more accurate QNM frequencies for different modes.
Hereafter we will  focus on the calculation of single-mode SNR.
For the complicated data, it is not easy to find the dominant mode in the gravitational perturbation.
Here we will use again the criteria suggested in \cite{Wang:2004bv} by examining min$\{\sqrt{\omega_R^2+\omega_I^2}\}$,
which tells us that the mode $n=0, l=2$ is dominant in both the non-singular and the Schwarzschild  BHs.
Taking into account that the $n=0,l=2$ mode always has the strongest amplitude $\mathcal{A}_{lm}$,
it gives us further confidence to employ the $n=0, l=2$ mode to calculate the SNR in our following discussion.

\begin{table}[!htbp]
\centering
\caption{QNMs frequency $M\omega$ for non-singular Bardeen  BHs.}
        \begin{tabular}{ccccccccccc}
    \hline\hline
    $l$ & $n$ & $\beta=0$& $\beta=0.3$ & $\beta=0.6$\\
    \hline
    $2$&$0$&$0.373675-0.088964i $&$0.406175 - 0.087325i $&$0.553008 - 0.094534i $&$ $\\
    $ $&$1$&$0.346827-0.273931i $&$0.381988 - 0.269701i $&$0.525271 - 0.291523i $&$ $\\
    $ $&$2$&$0.299998-0.478098i $&$0.383018 - 0.475991i $&$0.435513 - 0.485159i $&$ $\\
    $3$&$0$&$0.599443-0.092703i $&$0.626872 - 0.091561i $&$0.743159 - 0.089355i $&$ $\\
    $ $&$1$&$0.582644-0.281297i $&$0.612724 - 0.277584i $&$0.731708 - 0.268568i $&$ $\\
    $ $&$2$&$0.551686-0.479087i $&$0.586992 - 0.471962i $&$0.714495 - 0.445970i $&$ $\\
    $ $&$3$&$0.511943-0.690318i $&$0.692611 - 0.523967i $&$0.692336 - 0.618884i $&$ $\\
    $4$&$0$&$0.809178-0.094163i $&$0.835574 - 0.093074i $&$0.942654 - 0.088437i $&$ $\\
    $ $&$1$&$0.796631-0.284334i $&$0.824486 - 0.280904i $&$0.934782 - 0.266149i $&$ $\\
    $ $&$2$&$0.772709-0.479908i $&$0.803399 - 0.473636i $&$0.919857 - 0.446242i $&$ $\\
    $ $&$3$&$0.739836-0.683924i $&$0.774487 - 0.673961i $&$0.899410 - 0.630009i $&$ $\\
    $ $&$4$&$0.701514-0.898237i $&$0.740946 - 0.883478i $&$0.876439 - 0.818826i $&$ $\\
    $5$&$0$&$1.012295-0.094871i $&$1.039250 - 0.093792i $&$1.145670 - 0.088485i $&$ $\\
    $ $&$1$&$1.002221-0.285817i $&$1.030170 - 0.282482i $&$1.139280 - 0.266113i $&$ $\\
    $ $&$2$&$0.982695-0.480328i $&$1.012590 - 0.474430i $&$1.126900 - 0.445668i $&$ $\\
    $ $&$3$&$0.955004-0.680556i $&$0.987689 - 0.671579i $&$1.109250 - 0.628267i $&$ $\\
    $ $&$4$&$0.921081-0.888197i $&$0.957252 - 0.875455i $&$1.087490 - 0.814877i $&$ $\\
    $ $&$5$&$0.883335-1.104182i $&$0.923566 - 1.086900i $&$1.062670 - 1.005600i $&$ $\\
    \hline\hline
    \label{table3}
\end{tabular}
\end{table}

\subsection{SNR by LISA, TianQin and TaiJi}

Following the discussion in Section \ref{section3},
here we are going to discuss the SNR of the GW signal to be detected by  LISA for  non-singular Bardeen  BHs at first, and then we will make a comparison of SNR among different space GW detectors, such as LISA, TianQin and TaiJi.
For the non-singular Bardeen black hole and the Schwarzschild black hole having the same dominant mode,
it is easy to compare their single-mode SNR.

\begin{figure}[thbp]
\centering
\includegraphics[height=2.2in,width=3.2in]{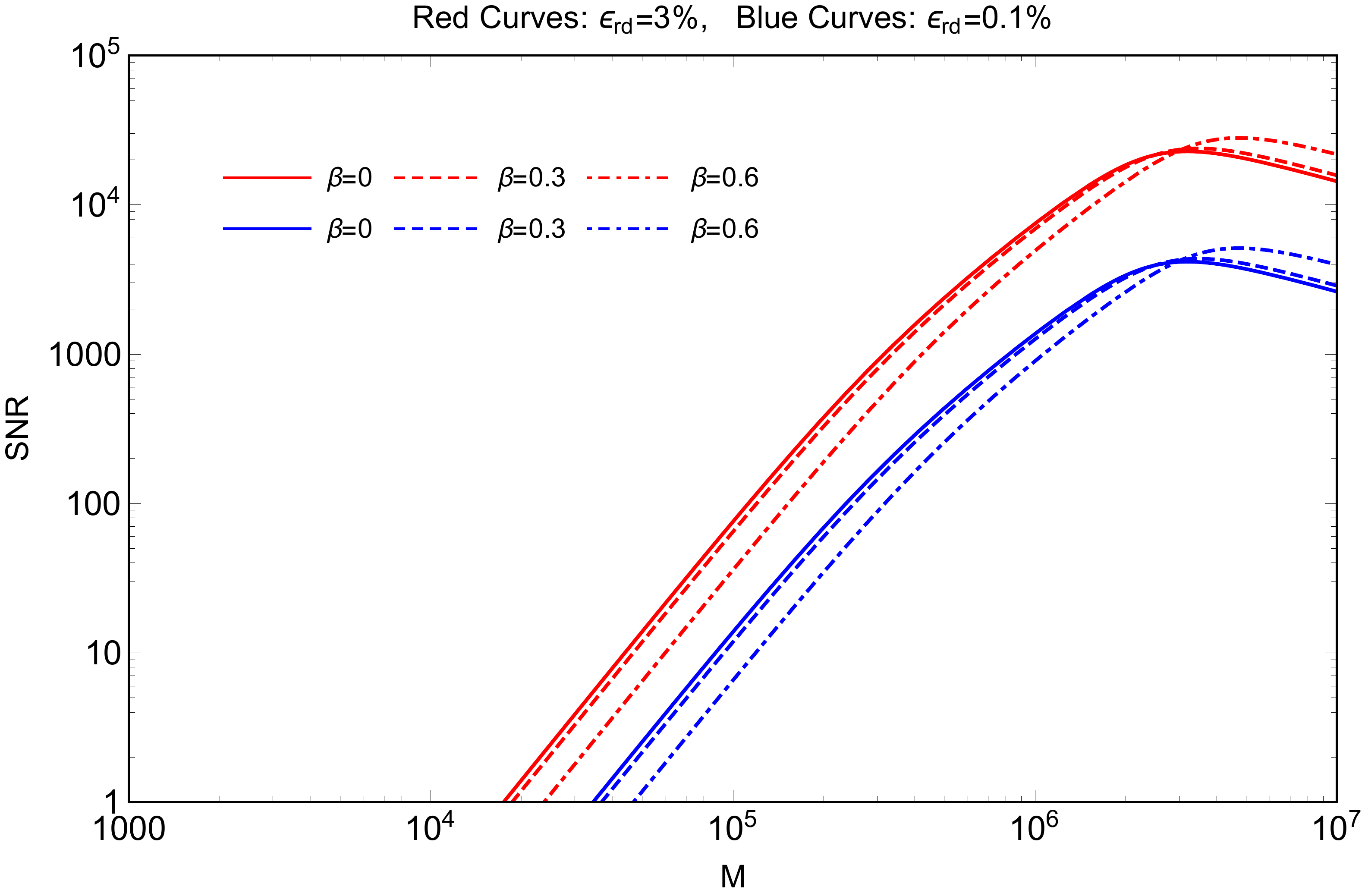}
\includegraphics[height=2.2in,width=3.2in]{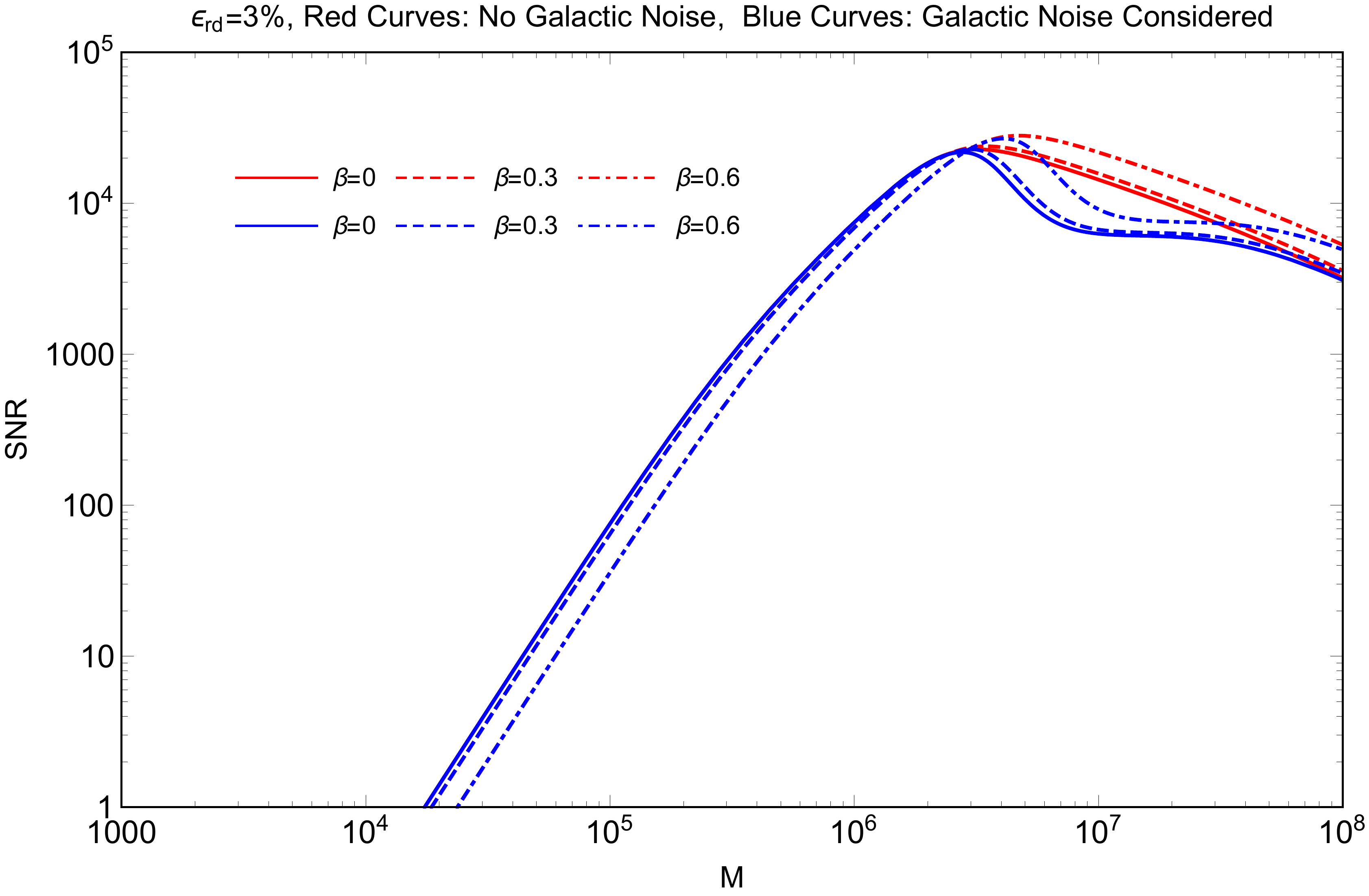}
\caption{The SNR behavior of LISA with the change of the black hole mass $M$ at a distance $D_L=3\,\mathrm{Gpc},z=0.54$ when  the angular index is taken $l=2$.  For the left plot, the red curves and blue curves correspond to radiation efficiency $\epsilon_{rd}=3\%$ and $0.1\%$, respectively. For the right plot, we set $\epsilon_{rd}=3\%$ and consider the comparison between SNR affected by the galactic noise (marked by blue lines) and SNR without including the galactic noise (marked by red curves). \label{fig5}}
\end{figure}

We show the SNR curves for the Bardeen  BHs with three different values,
$\beta=0,\beta=0.3$ and $\beta=0.6$ in Fig. \ref{fig5}.
In the left panel we do not consider the galactic noise,
while in the right panel the noise is included. The general property of the SNR in this case  is quite similar to that reported above for the non-singular conformal  BHs  in Section \ref{section3}.
When the black hole mass $M<2\times 10^6 M_{\odot}$,
the SNR for the Schwarzschild  BHs with $\beta=0$ is  higher than that of the non-singular Bardeen  BHs with a non-zero $\beta$.
However for more massive  BHs,
the SNR for the non-singular Bardeen black hole exhibits higher peaks for bigger $\beta$.
\begin{figure}[thbp]
\centering
\includegraphics[height=2.2in,width=3.2in]{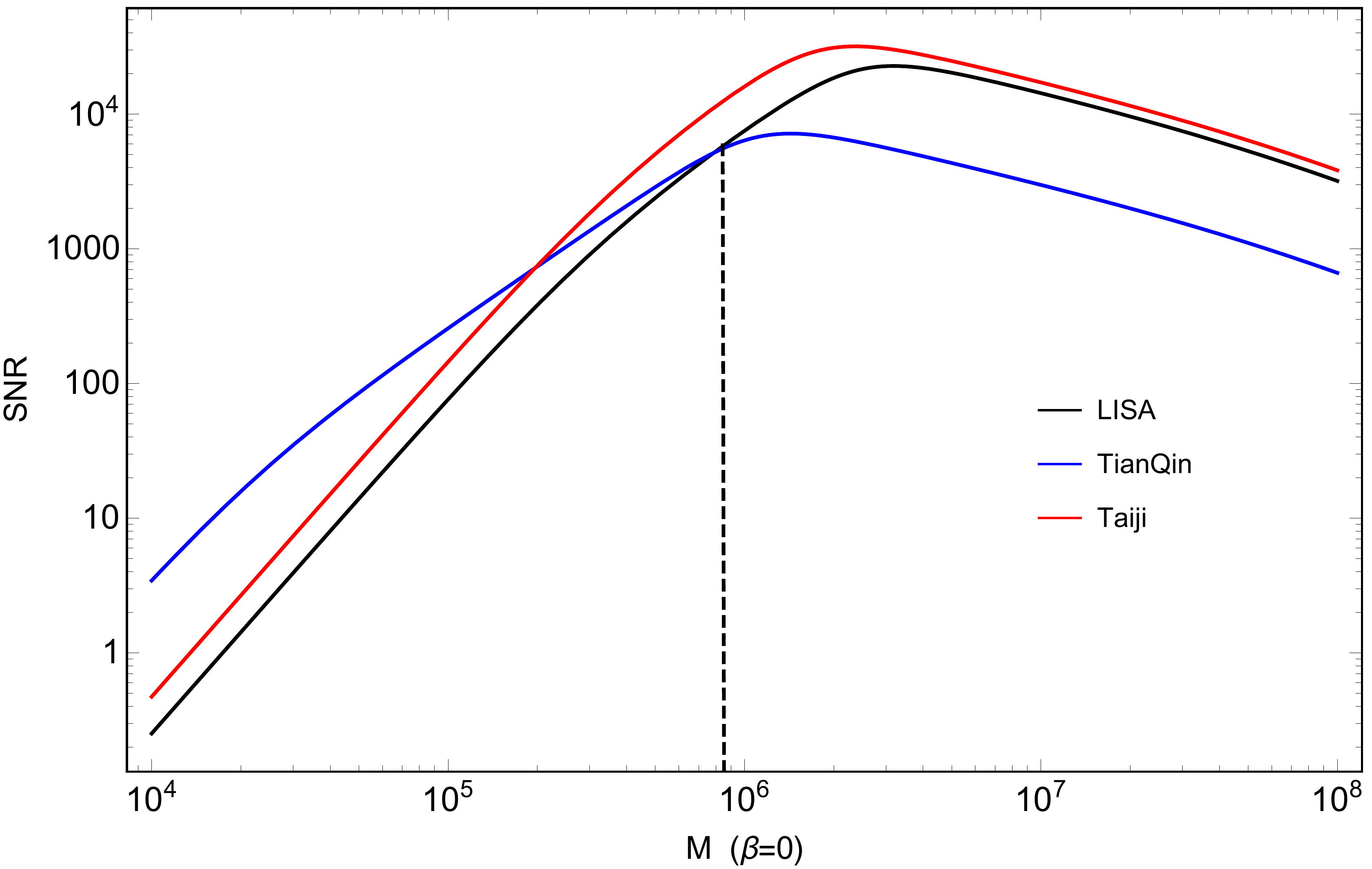}
\includegraphics[height=2.2in,width=3.2in]{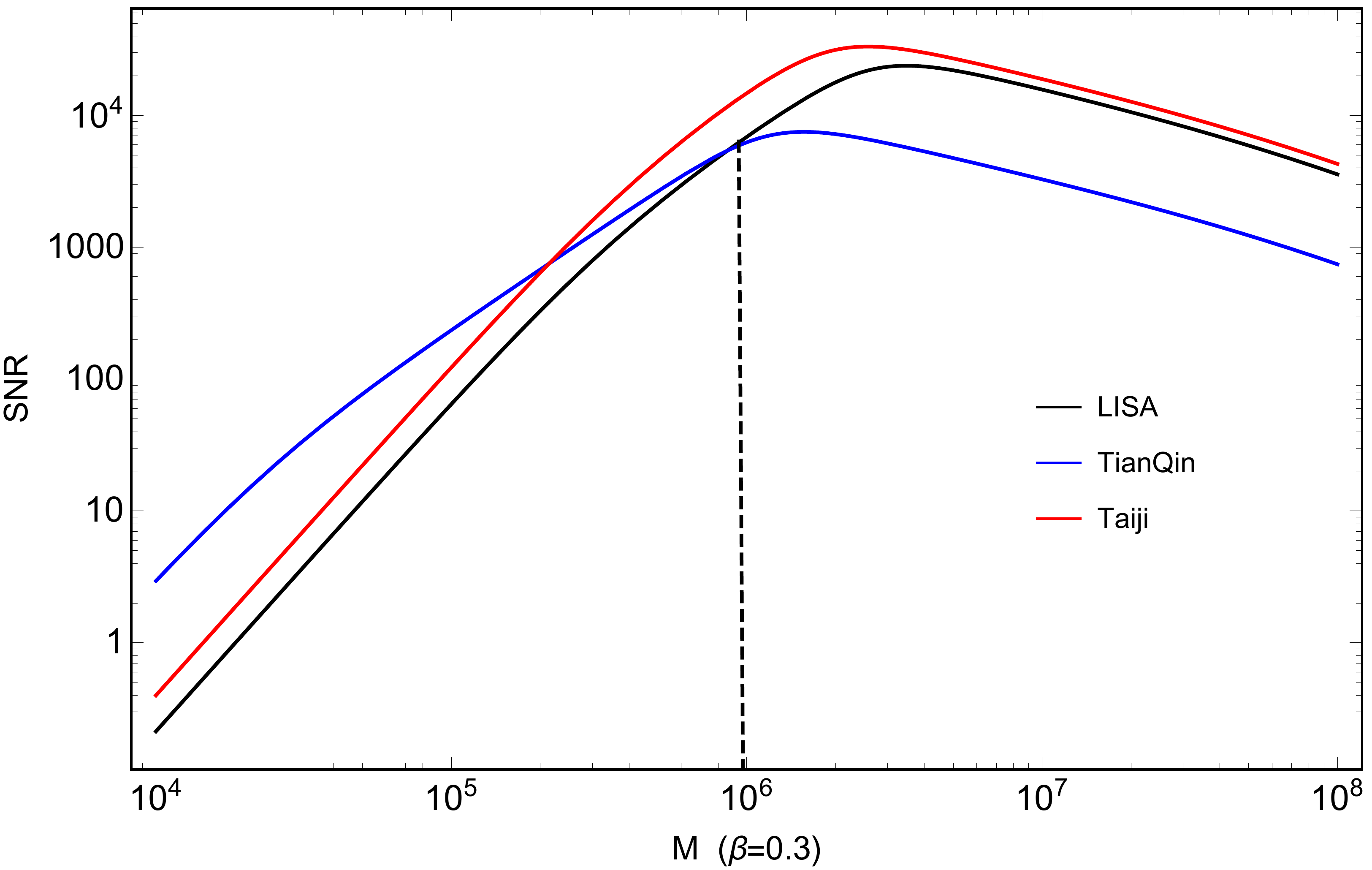}
\includegraphics[height=2.2in,width=3.2in]{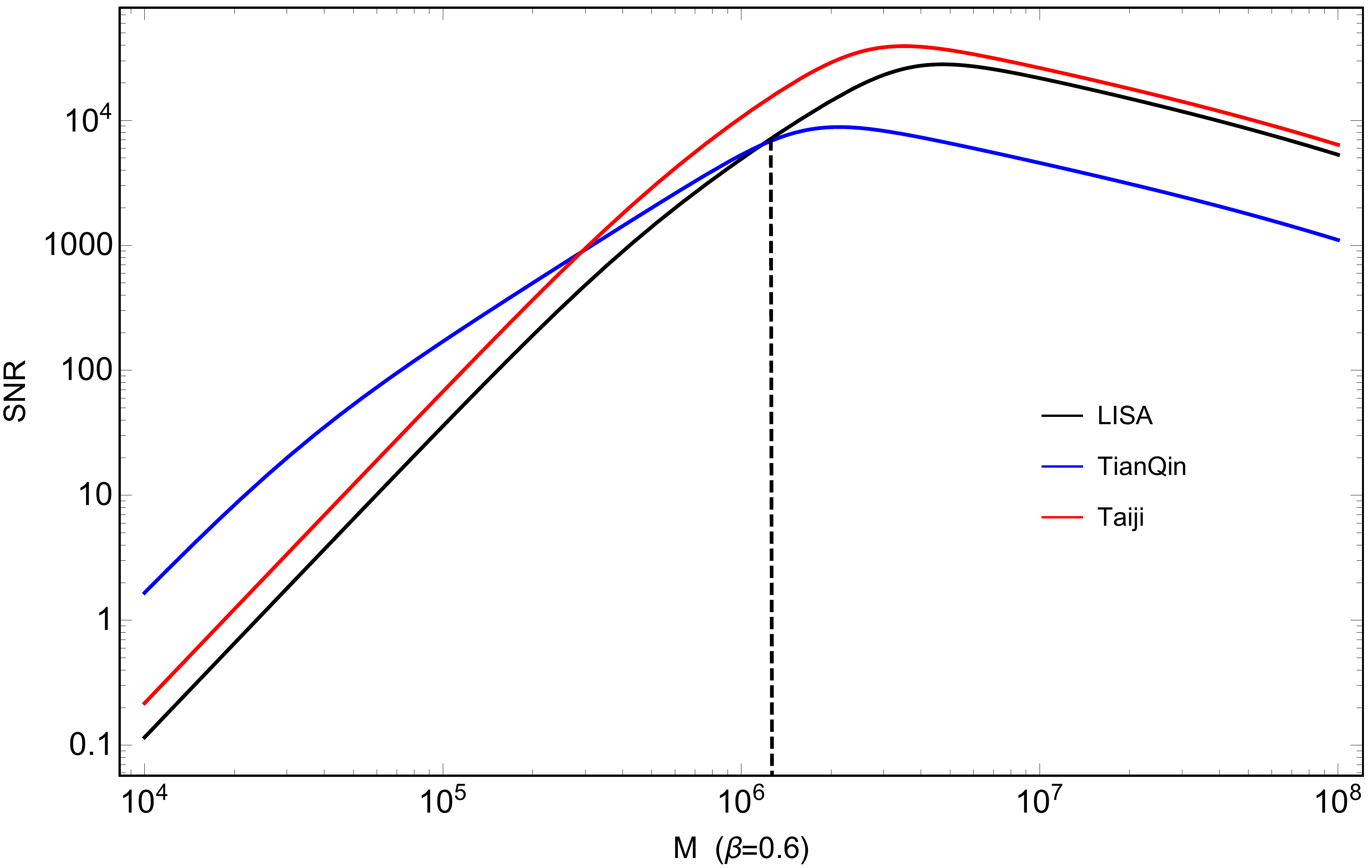}
\caption{The comparison of SNR among LISA and TianQin and TaiJi for the Bardeen  BHs.\label{fig6}}
\end{figure}
In Fig. \ref{fig6} we illustrate the comparison of SNR among LISA, TianQin and TaiJi.
The comparison shows that there exists a critical mass $M_{cri}$, below which TianQin is more sensitive to detect the GW signal,
while above this vaule LISA or TaiJi will detect the signal more sensitively.
This critical mass $M_{cri}$ increases with the increase of the Bardeen factor $\beta$.
In the whole frequency band, it is clear that TaiJi has higher SNR than LISA.
Again comparing the values of $M_{cri}$ and the locations of SNR peaks for different Bardeen factors,
LISA and TaiJi have more potential to distinguish  the Bardeen non-singular  BHs from the Schwarzschild ones.

\section{The uncertainty of parameter estimation}\label{section5}
In last two sections we have calculated the SNR by detection of single mode GWs sourced by  non-singular black holes. In this section, it is necessary to obtain the measurements errors of black holes parameters, such as mass $M$, conformal parameters $L, N$ and Bardeen parameter $\beta$. To this end, we are going to employ Fisher information matrix which is widely used to obtain the uncertainty in parameter estimation. In our calculation of Fisher matrix, we would like to follow the strategy presented in Ref. \cite{Cardoso}.
\subsection{Statistical Methods}\label{section5.1}
We define the inner product between two signals $h_1(t)$ and $h_2(t)$ by \cite{Cardoso}
\begin{equation}
(h_1|h_2)=2\int_{0}^{\infty}\frac{\tilde{h}_1^{\ast}\tilde{h}_2+\tilde{h}_2^{\ast}\tilde{h}_1}{S_N(f)}df,
\end{equation}
where the $S_N(f)$ is the noise spectral density for detetors, and $\tilde{h}_1(f)$ and $\tilde{h}_2(f)$ is the Fourier transform of the respective gravitational waveforms $h_1(t)$ and $h_2(t)$. With the definition of the inner product, the components of the Fisher matrix $\Gamma_{ab}$ are given by
\begin{equation}
\Gamma_{ab}=\left(\frac{\partial h}{\partial \theta^a}|\frac{\partial h}{\partial \theta^b}\right),
\end{equation}
where the $\theta$ are a set of parameters that the gravitational waveforms depend on. In the large SNR limit, if the noise is stationary and Gaussian, the best-fit parameters will have a Gauss distribution centered on the correct values \cite{Cutler:1997ta}. The probability that the GWs signal $s(t)$ is described by a set of given values of the source parameters $\theta^a$ is given by \cite{Cardoso}
\begin{equation}
p(\boldsymbol{\theta}|s)=p^{(0)}(\boldsymbol{\theta})e^{-\frac{1}{2}\Gamma_{ab}\delta\theta^a\delta\theta^b},	
\end{equation}
 where $\delta\theta^a=\theta^a-\hat{\theta}^a$ and $\hat{\theta}^a$ means the ``true" values of the parameters, $p^{(0)}(\boldsymbol{\theta})$ stands for the distribution of the prior information. The uncertainty in the measurement of parameter $\theta^a$ is represented by the rms error $\Delta\theta^a=(\langle(\delta\theta^a)^2\rangle)^{\frac{1}{2}}$ which can be calculated at large SNR by
 \begin{equation}
 \Delta\theta^a	\approx \sqrt{(\Gamma^{-1})^{aa}}.
 \end{equation}

The gravitational waveforms considered in our calculation are
\begin{subequations}\label{eq8}
\begin{align}
h_+(t)&=A^{+}e^{-\pi f_{lmn} t/Q_{lmn}}\cos[2\pi f_{lmn}t+\phi^{+}_{lmn}]S_{lmn},\\
h_{\times}(t)&=A^{+}N_{\times}e^{-\pi f_{lmn}t/Q_{lmn}}\sin[2\pi f_{lmn}t+\phi^{\times}_{lmn}]S_{lmn},
\end{align}
\end{subequations}
in which
\begin{equation}
A^{+}=\frac{M}{r}\mathcal{A}^+_{lmn}	, \quad A^{\times}=A^{+}N_{\times}=\frac{M}{r}\mathcal{A}^{\times}_{lmn},\quad \phi^{\times}_{lmn}=\phi^{+}_{lmn}+\phi^{0}_{lmn},
\end{equation}
where $N_{\times}$ is some numerical factor. For simplicity, we assume that we know $N_{\times}$ and $\phi^0_{lmn}$ such that the waveform only depends on four parameters $(A^{+},\phi^{+}_{lmn},f_{lmn},Q_{lmn})$. Specifically, in Kerr case the four parameters can also be represented by  $(A^{+},\phi^{+}_{lmn},M,j)$ because $f_{lmn}$ and $Q_{lmn}$ are only dependent on mass $M$ and angular momentum $j$. In black holes models found in alternative theories of gravity, the parameter basis could involve more parameters since we may need more parameters to describe $f_{lmn}$ and $Q_{lmn}$.

\begin{figure}[thbp]
\centering
\includegraphics[height=2.2in,width=3.2in]{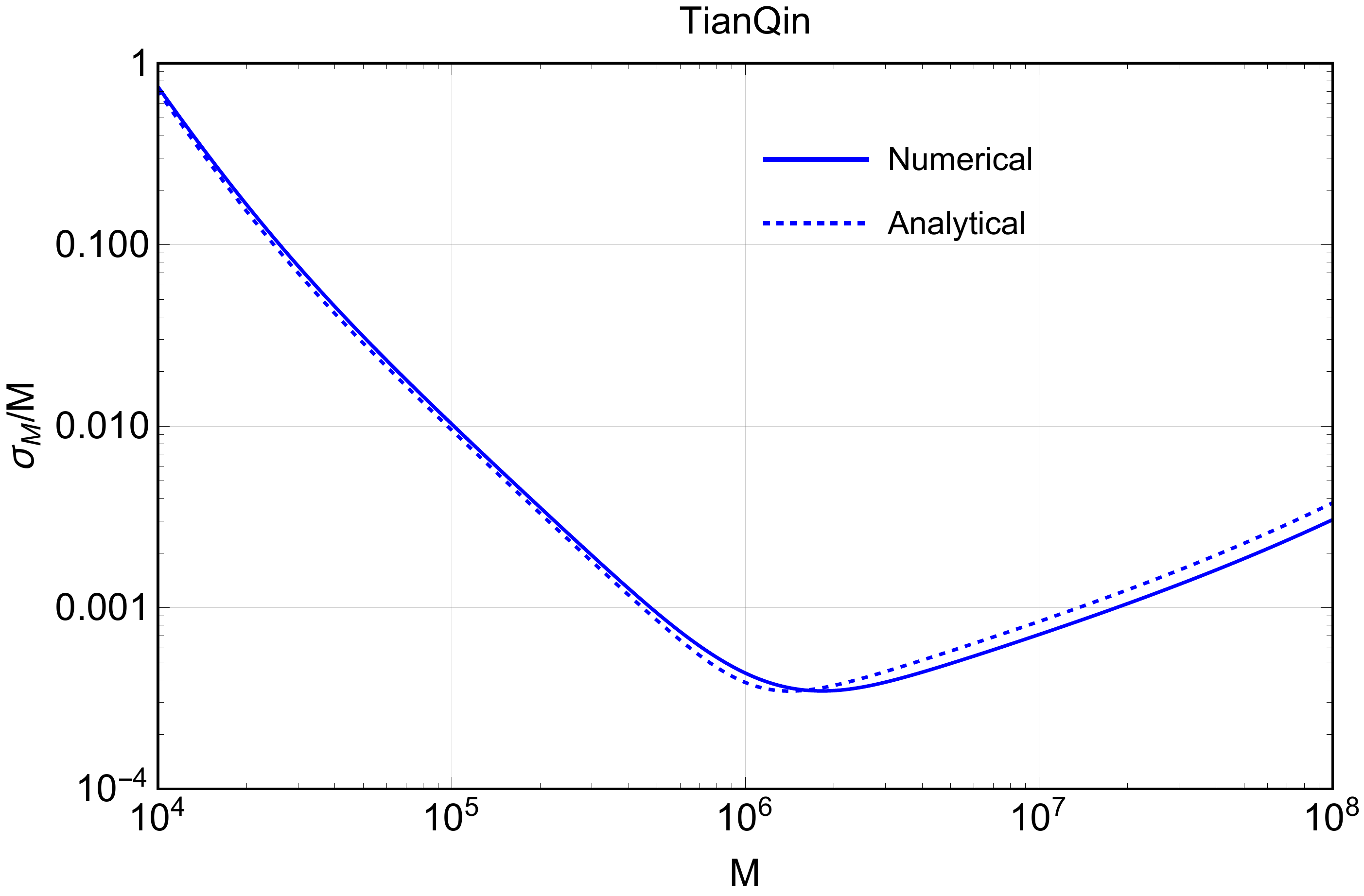}
\includegraphics[height=2.2in,width=3.2in]{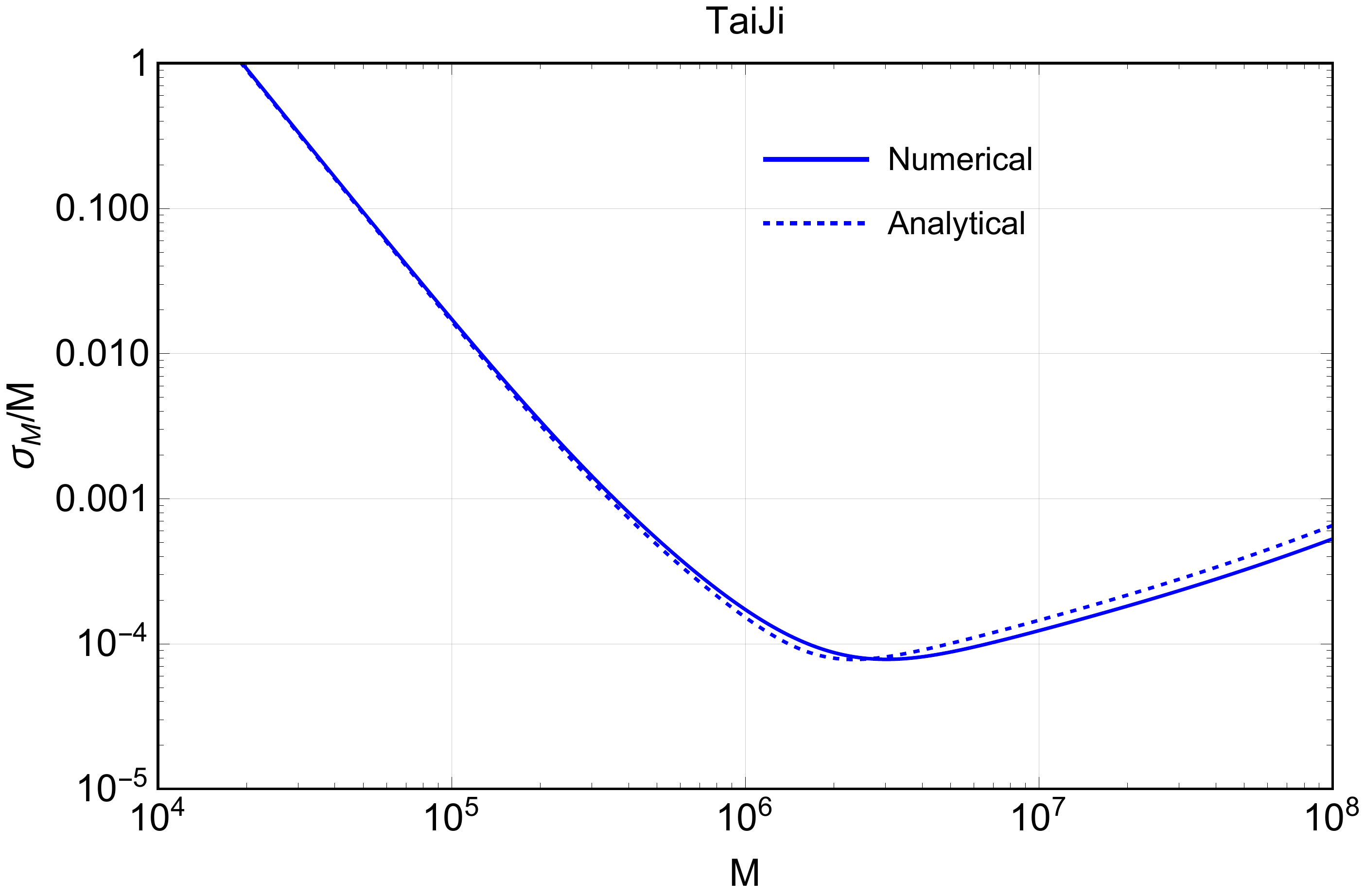}
\caption{The  mass error $\sigma_M/M$ for TianQin and TaiJi. The gravitational QNMs frequencies of Schwarzschild black holes used here are $l=2,n=0$ modes. The solid curves represent the numerical results and dotted curves stand for analytical results obtained by Eq. (\ref{eq7}). One can see that the numerical and analytical results are in good agreement with each other. In our calculation, we have assumed $N_{\times}=1, \phi^{\times}_{lmn}=\phi^{+}_{lmn}=0$, and $z=0.54, \epsilon_{ed}=3\%, D_L=3 \mathrm{Gpc}$. \label{fig10}}
\end{figure}

The parameter errors to leading order in $Q_{lmn}^{-1}$ in Kerr case have been analytically given in Ref. \cite{Cardoso}
\begin{subequations} \label{eq7}
\begin{align}
	&\sigma_{A^+}=\frac{\sqrt{2}A^+}{\rho_{FH}}\left|1+\frac{3\varpi}{8Q_{lmn}^2}\right|,\\
	&\sigma_{\phi_{lmn}^+}=\frac{1}{\rho_{FH}}\left|1- \frac{\varpi}{4Q_{lmn}^2}\right|,\\
	&\sigma_{M}=\frac{1}{\rho_{FH}}\left|\frac{2MQ_{lmn}f'_{lmn}}{f_{lmn}Q'_{lmn}}\left(1+\frac{1+4\varpi}{16Q_{lmn}^2}\right) \right|,\\
	&\sigma_{j}=\frac{1}{\rho_{FH}}\left|\frac{2Q_{lmn}}{Q'_{lmn}}\left(1+\frac{1+4\varpi}{16Q_{lmn}^2}\right) \right|,
\end{align}
\end{subequations}
where
\begin{equation}
f'_{lmn}=\frac{df_{lmn}}{dj},\quad Q'_{lmn}=\frac{dQ_{lmn}}{dj}, \quad \varpi=\frac{N_{\times}^2}{1+N_{\times}^2}\cos(2\phi^{\times}_{lmn})-\frac{1}{1+N_{\times}^2}\cos(2\phi^{+}_{lmn}).
\end{equation}
It is apparently to note that Eq. (\ref{eq7}) is also applicable to TianQin and TaiJi although these formulas are developed for LISA, since in Eq. (\ref{eq7}) the SNR formula $\rho_{FH}$ is the unique factor  related to characteristics of detector and this formula  is valid to TianQin and TaiJi as we have discussed previously. For completeness, we show the numerical and analytical results of the mass  errors   to demonstrate the validity of the Eq. (\ref{eq7}) for TianQin and TaiJi in Fig. \ref{fig10} from which we can see that the numerical results are in good agreement with  analytical results. In our calculation, as the choices made and explained in Ref. \cite{Cardoso}, we also have assumed $N_{\times}=1, \phi^{\times}_{lmn}=\phi^{+}_{lmn}=0$, and $z=0.54, \epsilon_{ed}=3\%, D_L=3 \mathrm{Gpc}$.

\subsection{Parameter Estimation Uncertainty of Non-Singular Black Holes}
In this section, we numerically calculate the errors in the measurements of parameters for the non-singular black holes under the condition $N_{\times}=1, \phi^{\times}_{lmn}=\phi^{+}_{lmn}=0$, and $z=0.54, \epsilon_{ed}=3\%, D_L=3 \mathrm{Gpc}$, as the condition we imposed in \ref{section5.1}. Note that the sensitivity curve of TaiJi behaves similarly to that of LISA, it is more interesting to compare the parameter detection errors between LISA and TianQin. The Bardeen black hole will be discussed firstly and then we move on to the discussion of conformal black hole.

\subsubsection{Bardeen Black Hole Case}
In Bardeen black hole case, we calculate Fisher matrix elements in  the parameter basis $(A^+,\phi^+_{lmn},M,\beta)$. However, mass $M$ and Bardeen parameter $\beta$ are not explicitly expressed in waveforms (\ref{eq8}) where $f_{lmn}$ and $Q_{lmn}$ are present and they are dependent on the black hole parameter $M$ and $\beta$. The relations between $f_{lmn}, Q_{lmn}$ and $M, \beta$ can not be analytically  given. We overcome this problem by relating QNMs frequencies obtained numerically  to black holes parameters with fitting formulas
\begin{subequations}
\begin{align}
f_{lmn}&=\frac{\omega_{lmn}}{2\pi}=\frac{\mathcal{F}_{lmn}}{2\pi (1+z) M}=\frac{f_1+f_2(\beta_{max}-\beta)^{f_3}}{2\pi (1+z)M},\\
Q_{lmn}&=q_1+q_2(\beta_{max}-\beta)^{q_3},
\end{align}	
\end{subequations}
where $\beta_{max}=\frac{4}{3\sqrt{3}}$ is the maximum value of $\beta$, and
\begin{subequations}\label{eq9}
\begin{align}
f_1&=0.254409, \quad f_2=0.094561, \quad f_3=-0.656729,\\
q_1&=-0.238375,\quad q_2=2.187814, \quad q_3=-0.208734.
\end{align}	
\end{subequations}
After substituting Eq. (\ref{eq9}) to waveforms (\ref{eq8}), then the Fisher matrix can be obtained  in the parameter basis $(A^+,\phi^+_{lmn},M,\beta)$. Actually, we have used similar method in Section \ref{section5.1} to estimate parameter errors  and the fitting formulas and fitting coefficients are provided in Ref. \cite{Cardoso}.

\begin{figure}[thbp]
\centering
\includegraphics[height=2.2in,width=3.2in]{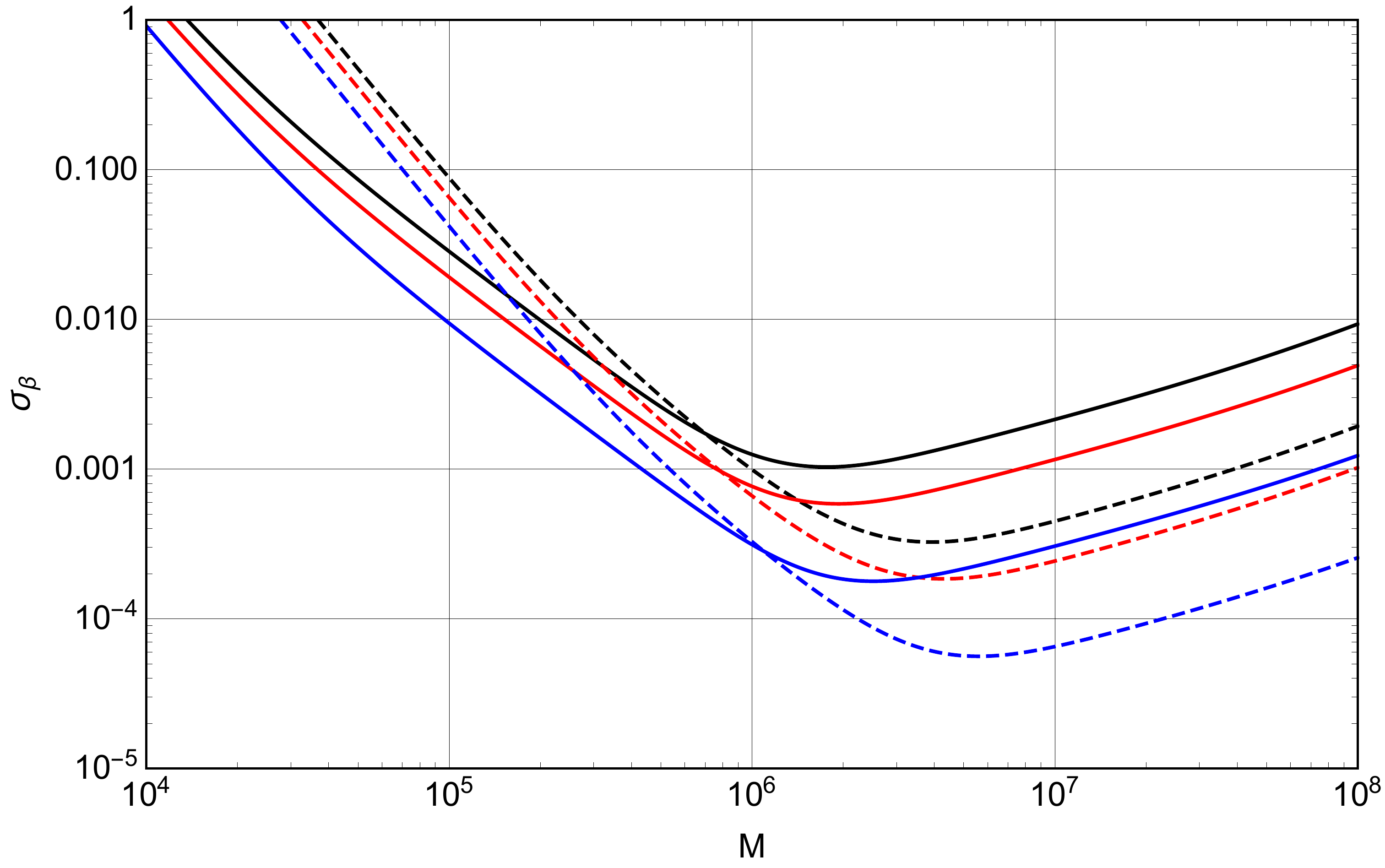}
\includegraphics[height=2.2in,width=3.2in]{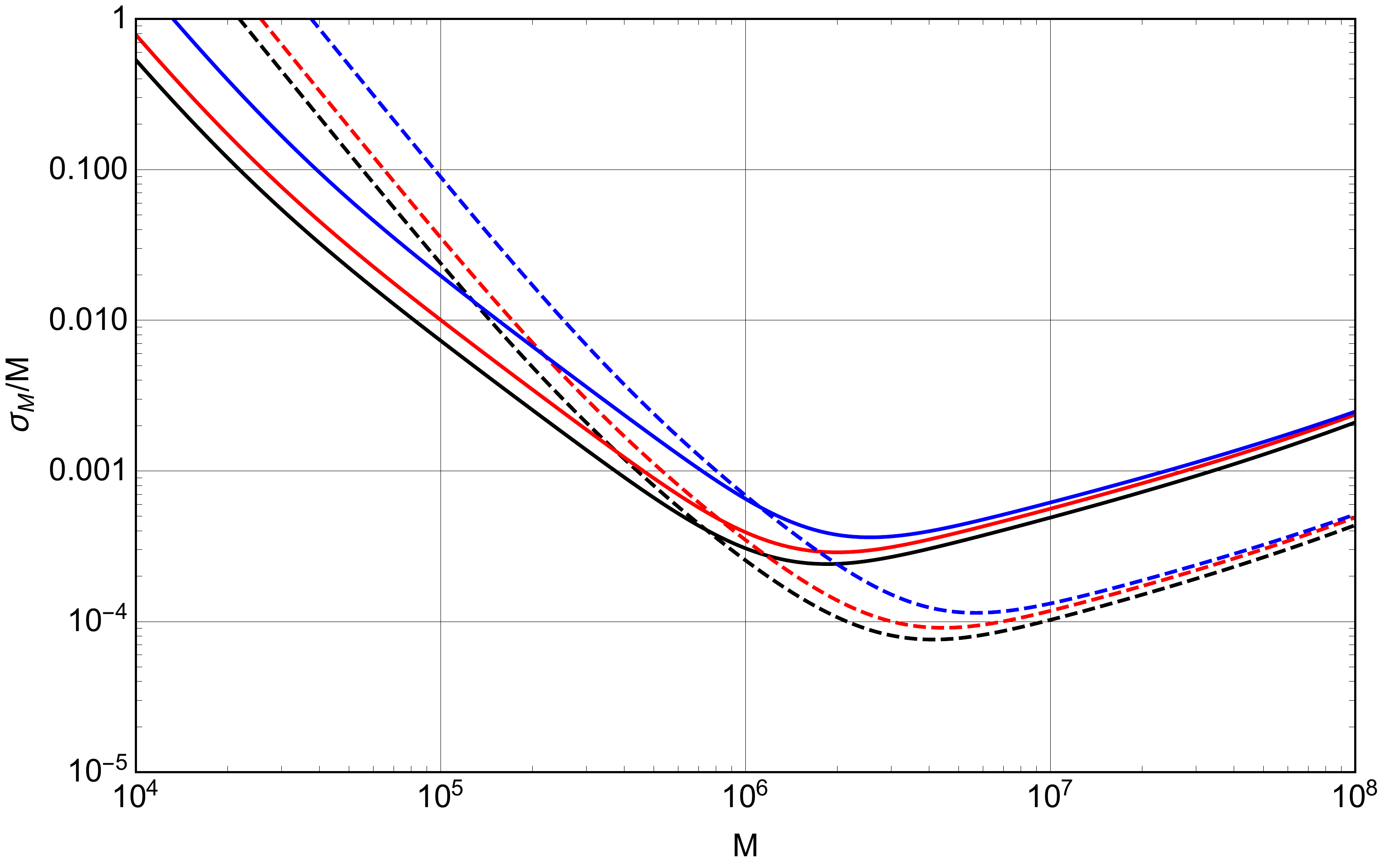}
\caption{The dependence of parameter estimation errors for Bardeen parameter $\beta$ (left panel) and black hole mass $M$ (right panel) on the mass $M$. The black, red and blue curves correspond to $\beta=0,\beta=0.3$ and $\beta=0.6$, respectively. The dashed lines represent the parameter errors in the GWs detection by LISA and solid lines represent TianQin. \label{fig11}}
\end{figure}

In Fig. \ref{fig11} we show the dependence of parameter estimation errors for Bardeen parameter $\beta$ (denoted by $\sigma_{\beta}$) and black hole mass $M$ (denoted by $\sigma_{M}/M$) on the mass $M$ in the GWs detection by LISA (dashed lines) and TianQin (solid lines). The $\sigma_{\beta}$ curves are shown in the left panel from which we can see that with the increase of $\beta$ the errors $\sigma_{\beta}$ decrease. For both detectors, the curves behave in the same way as $\sigma_{\beta}$ decrease with the increase of $M$ until to some critical mass and then increase with the further increase of $M$, and the critical mass for TianQin is about $ 2\times 10^6 M_{\odot}$ while for LISA it is about $4 \times 10^6 M_{\odot}$. In the higher frequency region (corresponding to $M\lesssim 10^6 M_{\odot}$), $\sigma_{\beta}$ by LISA is bigger than that of TianQin, and the opposite result can be observed in the lower frequency band ($M\gtrsim 10^6 M_{\odot}$). That is to say  TianQin can detect parameter $\beta$ more precisely than LISA for relatively smaller black holes, while for more massive black holes, LISA can make a more precision detection. In the right panel, we show the mass error $\sigma_{M}/M$ which behaves generally similar to $\sigma_{\beta}$ but with two two different properties. The higher value of $\beta$ is related to a bigger mass error which is opposite to the behavior of $\sigma_{\beta}$. On the other hand, the differences of the value of $\sigma_{M}/M$ arising from the different $\beta$ will be reduced by increasing black hole mass. One can find that the  errors in both panel become unacceptably large for mass $M\lesssim 10^5M_{\odot}$, but in general we can expect good detection accuracies for black hole mass $M\gtrsim 3\times 10^5 M_{\odot}$ beyond which  the errors are smaller than $0.01$. The best precision detections are provided by LISA for the black hole mass $M\approx 5 \times 10^6M_{\odot}$ with both parameter errors $\sigma_{\beta}$ (when $\beta=0.6$) and $\sigma_{M}/M$ (when $\beta=0$) smaller than $10^{-4}$. Further more, we note that the characteristics of the behaviors of the errors curves can be reflected by the property of SNR $\rho_{FH}$, as the factor  $\rho_{FH}^{-1}$ is present in  the  analytical expressions of errors in Eq. (\ref{eq7}).

\subsubsection{Conformal Black Hole Case}

It was assumed  in Ref. \cite{Toshmatov:2017kmw} that the most natural non-singular black hole candidates from the numerous conformally invariant solutions  are those which have less violation of the energy conditions. On the other hand, the QNMs frequencies can be approximated analytically when $N$ is large enough, and hence the fitting formulas relating black hole parameters to QNMs are not required. Therefore, it is necessary and interesting to estimate errors in parameters measurements  at large $N$ limit. For the parameter space, we should point out that it would be more rigorous  to include $N$ in the parameter space. However, as the less violation of the energy conditions for large $N$ \cite{Toshmatov:2017kmw} but no limitations to parameter $L$ are implemented, we are more interested in exploring parameter $L$ and hence assuming a large value of $N$ is known. As a result, we do our calculation on the  basis of parameter space $(A^+,\phi^+_{lmn},M,L)$ which helps simplify our computation, and $N=100$ is assumed.

The effective potential Eq. (\ref{eq10}) can also be written as \cite{Chen:2019iuo}
\begin{equation}
	V(r)=f(r)\left(\frac{l(l+1)}{r^2}-\frac{3}{r^3}+F_1(r)N+F_2(r)N^2\right),
\end{equation}
where
\begin{align}
F_1(r)&=-\frac{2L^2(5r^3-6r^2+3L^2 r-4L^2)}{r^3(r^2+L^2)^2}\\
F_2(r)&=\frac{4L^4(r-1)}{r^3(r^2+L^2)^2}.
\end{align}
In large $N$ limit, the effective potential can be approximated as
\begin{equation}
V(r)\approx f(r)F_2(r)N^2.
\end{equation}
The approximation form of the potential suggests that we can use WKB approach  to work out QNMs frequencies. In the calculation of QNMs frequencies, the 6th order WKB formula is given by \cite{Schutz,PhysRevD.35.3621,Konoplya:2003ii}
\begin{equation}
\frac{i(\omega^2-V_m)}{\sqrt{-2V''_m}}-\sum_{i=2}^{6}\Lambda_i=n+\frac{1}{2},
\end{equation}
where $m$ represents the quantities evaluated at the maximum (peak)of the potential, and we denote the location of the peak of potential as $r_m$ which can be obtained analytically. $V''_{m}$ denotes the value of the second order derivative of potential with respect to $r_{\ast}$ calculated at $r_m$. $\Lambda_i$ are higher order correction terms which can be ignored at large $N$ (or large $l$) limit and only the 1st order term is left, as in our current consideration. Based on 1st order WKB formula, the QNMs frequencies are given by
\begin{equation}
\omega\approx \sqrt{V_m}-i\left(n+\frac{1}{2}\right)\sqrt{-\frac{1}{2}\frac{V''_{m}}{V_m}}=N\sqrt{(fF_2)_m}-i\left(n+\frac{1}{2}\right)\sqrt{-\frac{1}{2}\frac{(fF_2)''_m}{(fF_2)_m}}.
\end{equation}
With the help of this expression of QNMS frequencies, we can get numerical elements of Fisher matrix and hence obtain parameter detection errors.

\begin{figure}[thbp]
\centering
\includegraphics[height=2.2in,width=3.2in]{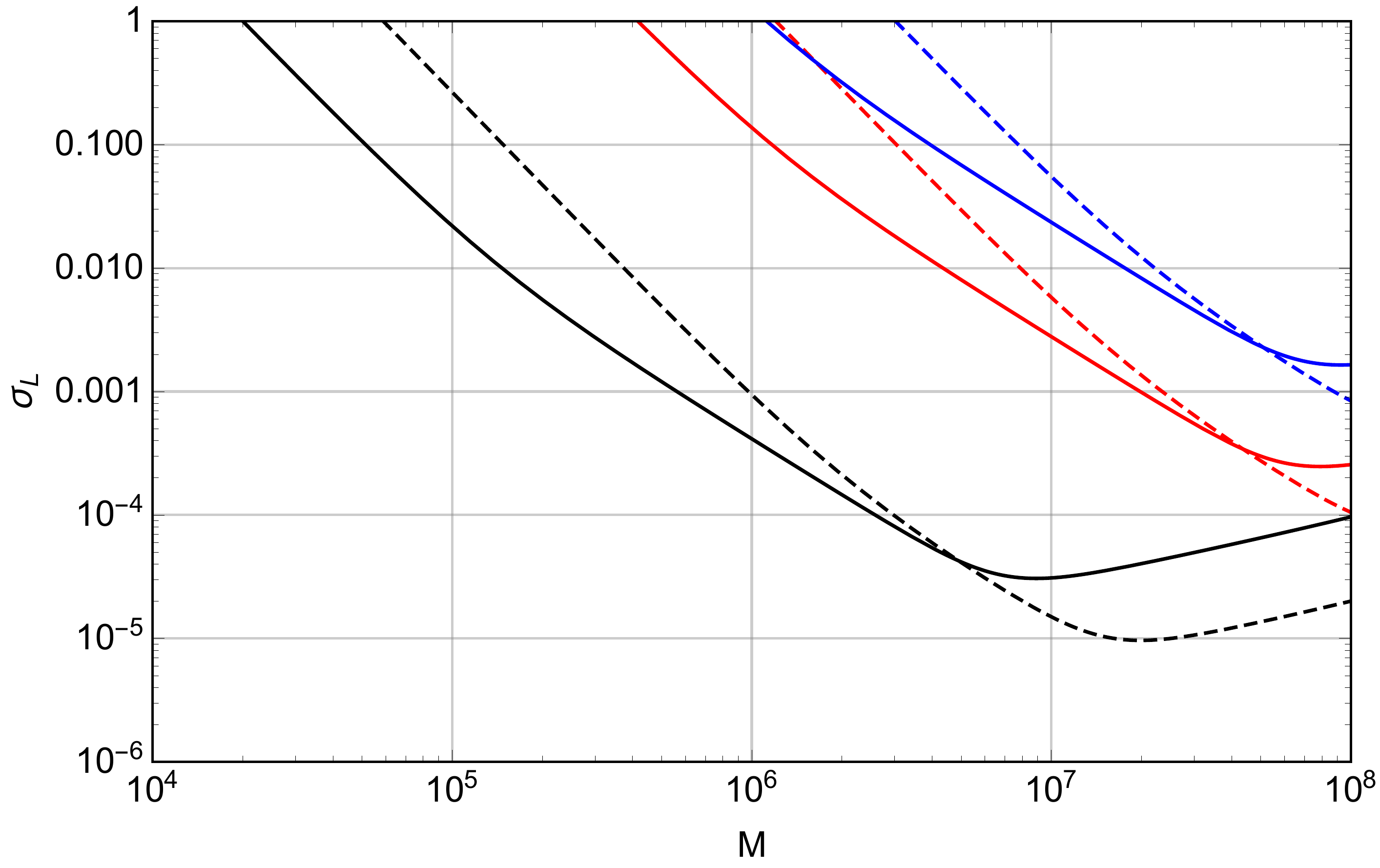}
\includegraphics[height=2.2in,width=3.2in]{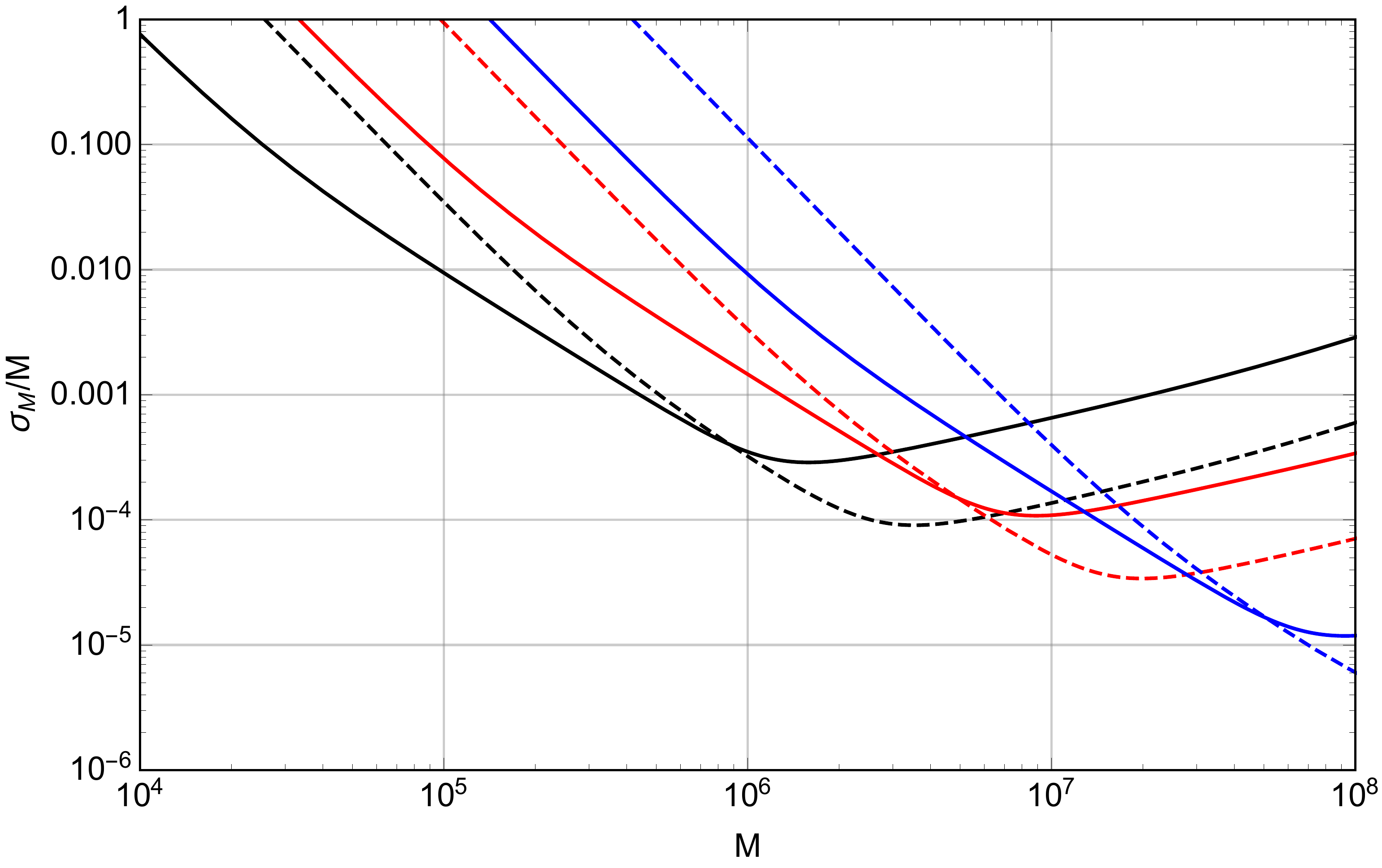}
\caption{The behavior of parameter errors under the change of black hole mass. For both panels, solid lines correspond to errors by TianQin and dashed lines correspond to LISA. The left plot shows the behavior of $\sigma_L$ standing for the errors of parameter $L$ under different $L$ value, the black, red and blue lines correspond to $L=0.5, L=4$ and $L=10$, respectively. The right plot shows the errors of mass, in this case we take $L=0.2, L=0.5$ and $L=10$ which correspond to black, red and blue lines respectively.\label{fig12}}
\end{figure}

We show our results in Fig. \ref{fig12} where the solid lines represent errors by TianQin and the dashed lines by LISA. The left panel shows the errors of parameter $L$. As expected,  TianQin can make more precision detection of parameter in higher frequency band while for lower frequency region LISA is more sensitive. For a small value of $L=0.5$ (indicated by black lines), the value of $\sigma_L$ can be as low as $10^{-5}$ with black hole mass around $M\approx 10^7 M_{\odot}$ suggesting that we can expect a great accuracy. While when we increase the value of $L$ to $L=4$ (red lines) and $L=10$ (blue lines), the value of $\sigma_L$ grows which means that the  precision becomes worse. Therefore a smaller true value of $L$ will provide us a more accurate measurement. On the right panel, we show the detection errors of mass parameter. To get an accurate enough measurement of mass $M$ (say, errors smaller than $10^{-2}$), the critical mass value to reach this accuracy will become larger when increasing $L$. However, a larger $L$ can lead to a smaller minimum of errors at some certain large mass.

\section{Conclusions}

In this paper we have calculated the quasinormal modes (QNMs) of the non-singular Bardeen  BHs
and singularity-free  BHs in conformal gravity.
We have also calculated their corresponding SNR  in the single-mode waveform detection of GWs by
the future space based interferometers,
such as  LISA, TianQin and TaiJi. We have found that the approximate formula (\ref{eq4}) of SNR is not only valid for LISA,
but also applicable to  TianQin and TaiJi.
Using this approach, we have calculated SNR for LISA, TianQin and TaiJi.
We have investigated the impact of the conformal factor on the behavior of the SNR and found that the increase of the conformal factor will lead to a higher SNR for more massive  BHs.  For the Bardeen  BHs,
similar phenomena are also observed that a bigger Bardeen parameter $\beta$ will result in a higher SNR,
when the Bardeen black hole is very massive. Once the black hole mass is $\sim10^6M_{\odot}$, usually the GW perturbation in the Schwarzschild black hole has higher SNR.
The signature of the non-singular modification will emerge when  BHs become more massive.
Comparing the SNRs among LISA, TianQin and TaiJi,
we found that the SNR of TianQin is always higher than that of LISA and TaiJi when the black hole is not so massive.
However for the black hole with mass over a critical mass,
LISA and TaiJi will have stronger SNR compared to that of TianQin.
Interestingly, this critical mass increases when the black hole deviates significantly from the Schwarzschild black hole.
In our study, we have found that the effect of the galactic confusion noise is not negligible,
and  its influence on the Schwarzschild black hole appears when the black hole mass is a few times of $10^6M_{\odot}$,
but for non-singular  BHs the effect of the galactic noise will play an important role only for more massive holes.
For the non-singular  BHs, considering that their SNR peaks and dips appear for more massive  BHs,
it is expected that the LISA and TaiJi have more potential to distinguish them from the Schwarzschild black hole.

We have investigated the errors in parameter estimation for non-singular black holes by TianQin and LISA, and find that the analytical formulas of errors developed in Ref. \cite{Cardoso} for LISA can also be applied to TianQin and TaiJi. As expected, TianQin can make more precise detection than LISA for smaller black holes (higher frequency), while in larger black holes regime (lower frequency) the more accurate measurements of parameters are provided by LISA. In the detection of Bardeen parameter $\beta$, we find that a  higher true value of $\beta$ will result in a more precision detection, but the mass detection accuracy  will become worse. In the detection of parameter $L$ for conformal black holes, it is found that a smaller value of $L$ is favored in the sense that higher values of $L$ correspond to less accurate detection. In general, in the detection of parameters for both non-singular black holes, our results suggest that we can expect good accuracy in the future GWs detections by TianQin and LISA, as well as TaiJi. Therefore, it is promising to explore the non-singular black holes with the future space-based detectors.

We have only studied the SNR for the single mode detection in this paper, and it is worth extending the discussion to multi-mode detections and making parameter estimation (rather than just estimation errors) with the detected ringdown signals.  For the multi-mode discussion, we have provided very accurate QNM frequency samples, which contain important properties for non-singular BHs. Besides we have only concentrated on  BHs without angular momenta. Considering that  BHs with rotation are more realistic in the universe, so it would be very interesting   to generalize our investigations to probe QNMs and SNR for rotating non-singular  BHs.

\begin{acknowledgments}
This research was supported in part by the National Natural Science
Foundation of China under Grant No. 11675145, 12075202 and 11975203, and
the Major Program of the National Natural Science Foundation of China under Grant No. 11690021.
\end{acknowledgments}

\appendix

\section{The approximation  formula of SNR for TianQin and TaiJi}
Although the approximate formula of SNR given by Eq. (\ref{eq4}) from Ref. \cite{Cardoso} was mainly developed in the context of LISA  and the process of deriving Eq. (\ref{eq4}) is dependent on the detector characteristics, the authors in Ref. \cite{Cardoso} claimed that the expressions used in the derivation are valid for any interferometric detectors. To prove this claim, we calculate the SNR  by using Eq. (\ref{eq4}) for four QNMs in the Kerr  BHs with $(l,m)=\{(2,1),(2,2),(3,3),(4,4)\}$ considered in Ref. \cite{Shi},  and compare the SNR of TianQin and TaiJi calculated by doing the full integral with the formula of SNR, given by Eq. (\ref{eq3}).

\begin{table}[!htbp]
\centering
\caption{Fitting coefficients given by Ref. \cite{Cardoso}}
        \begin{tabular}{ccccccccccc}
    \hline\hline
    $(l,m)$ & $f_1(l,m)$ & $f_2(l,m)$& $f_3(l,m)$ & $q_1(l,m)$ & $q_2(l,m)$& $q_3(l,m)$\\
    \hline
    $(2,1)$ & $0.6000$ & $-0.2339 $ & $0.4175 $ & $-0.3000$ &  $2.3561$ &$-0.2277$\\
    $(2,2)$ & $1.5251$ & $-1.1568 $ & $0.1292 $ & $0.7000 $ &  $1.4187$ &$-0.4990$\\
    $(3,3)$ & $1.8956$ & $-1.3043 $ & $0.1818 $ & $0.9000 $ &  $2.3430$ &$-0.4810$\\
    $(4,4)$ & $2.3000$ & $-1.5056 $ & $0.2244 $ & $1.1929 $ &  $3.1191$ &$-0.4825$\\
    \hline\hline
    \label{table4}
\end{tabular}
\end{table}

\begin{figure}[thbp]
\centering
\includegraphics[height=2.2in,width=3.2in]{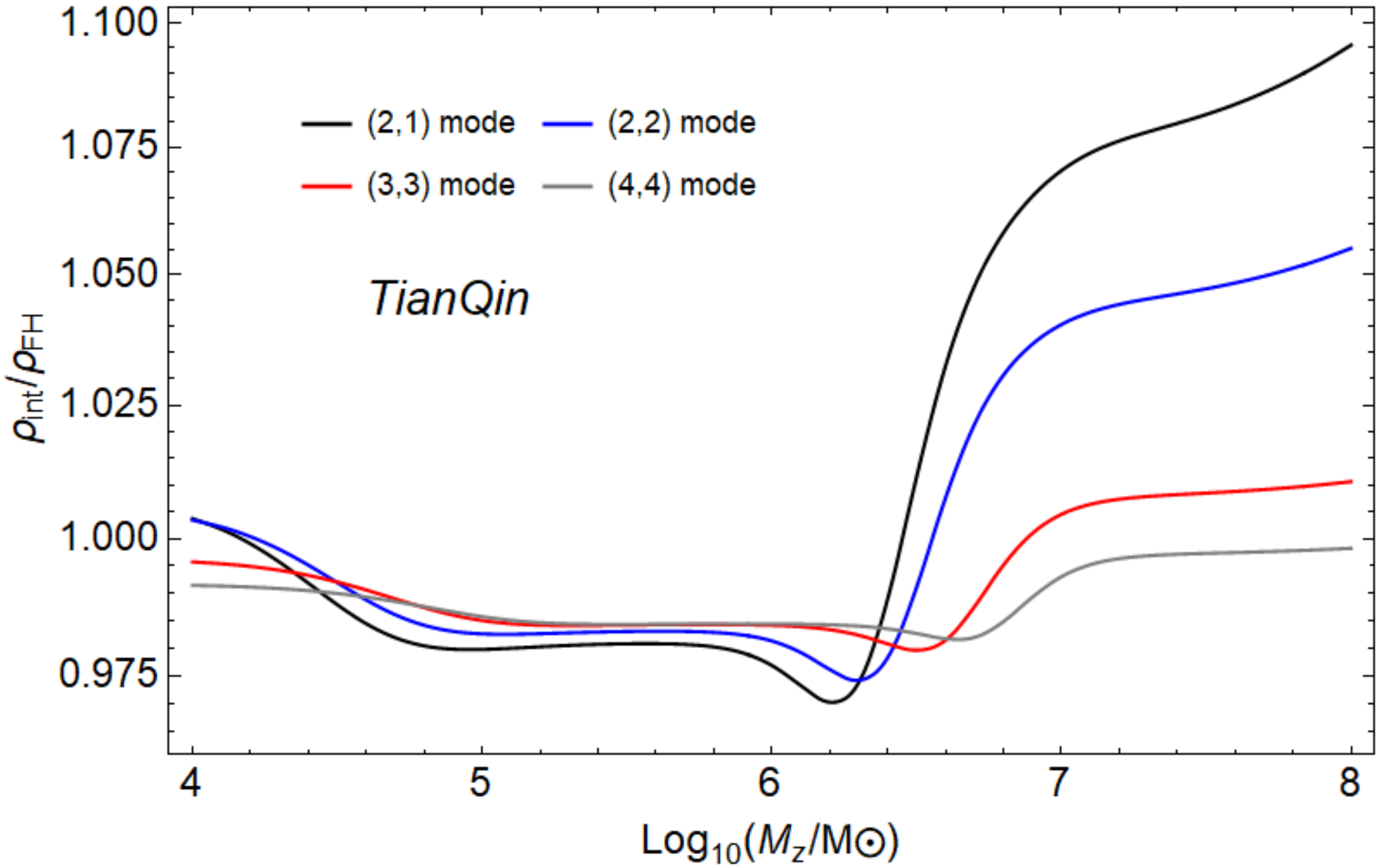}
\includegraphics[height=2.2in,width=3.2in]{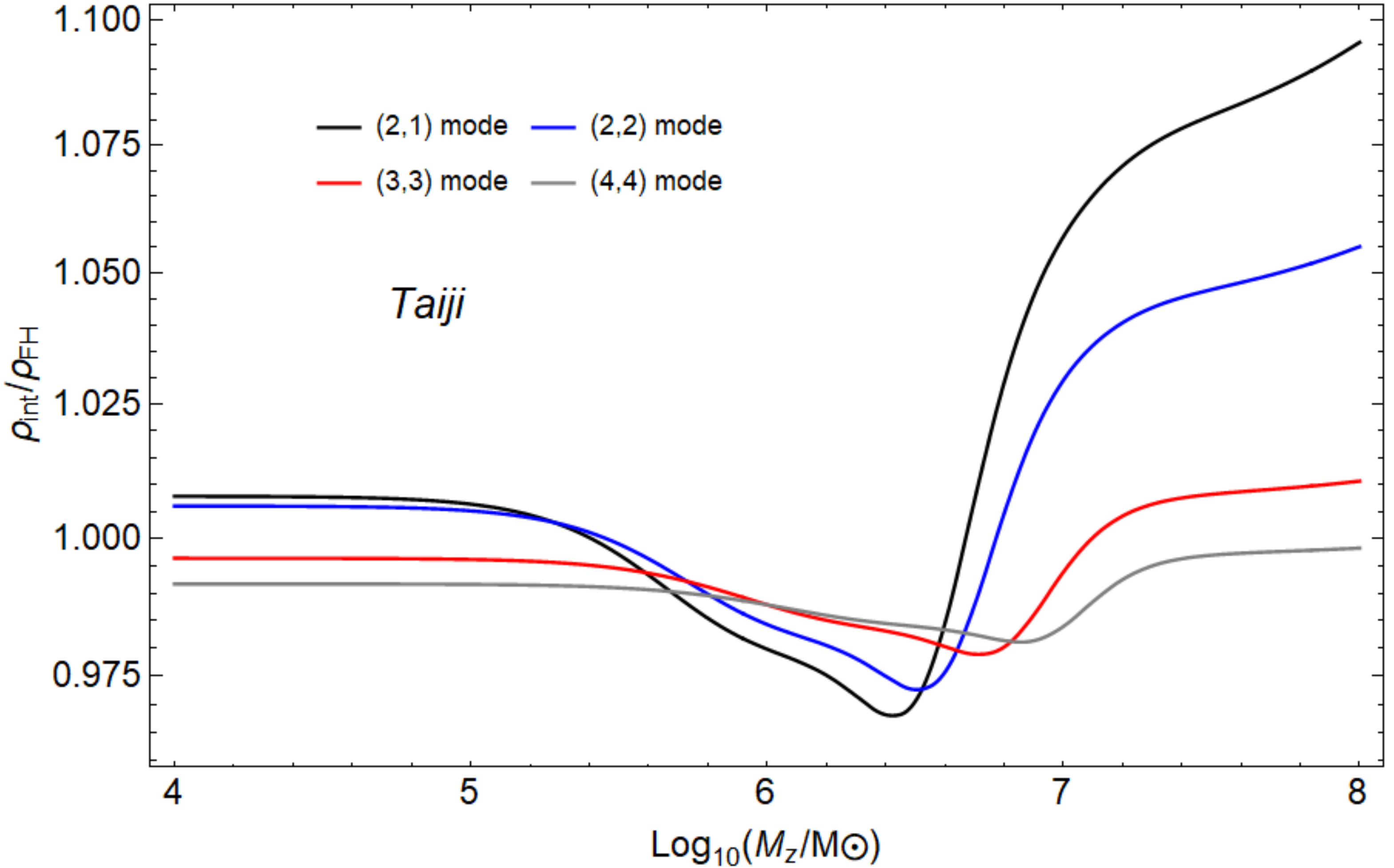}
\caption{The SNR ratio of $\rho_{\mathrm{int}}$ to $\rho_{FH}$. In this plot we take $\chi_{f}=0.76$, $\chi_{eff}=0.3$, $\nu=2/9$ and luminosity distance $
r=15\mathrm{Gpc}$.\label{fig9}}
\end{figure}

The fitting formulae of the oscillation frequency $\omega_{lm}$ and damping time $\tau_{lm}$ for the remnant Kerr black hole with  the redshifted mass $M_z=(1+z)M$ are given by \cite{Cardoso}
\begin{subequations}
\begin{align}
\omega_{lm}&=\frac{f_1(l,m)+f_2(l,m)(1-\chi_f)^{f_3(l,m)}}{M_z},\\
\tau_{lm}&=\frac{2(q_1(l,m)+q_2(l,m)(1-\chi_f)^{q_3(l,m)}}{\omega_{lm}},
\end{align}
\end{subequations}
where $\chi_f$ is the final spin parameter and the fitting coefficients are listed in Table. \ref{table4}. The amplitudes $\mathcal{A}_{lm}^{+}=\mathcal{A}_{lm}^{\times}=\mathcal{A}_{lm}$ are given by \cite{Kamaretsos,Meidam}
\begin{subequations}
\begin{align}
\mathcal{A}_{22}(\nu)&=0.864\nu,\\
\mathcal{A}_{21}(\nu)&=0.43(\sqrt{1-4\nu}-\chi_{eff})\mathcal{A}_{22}(\nu),\\
\mathcal{A}_{33}(\nu)&=0.44(1-4\nu)^{0.45}\mathcal{A}_{22}(\nu),\\
\mathcal{A}_{44}(\nu)&=(5.4(\nu-0.22)^2+0.04)\mathcal{A}_{22}(\nu),
\end{align}
\end{subequations}
and
\begin{align}
&\nu=\frac{m_1m_2}{(m_1+m_2)^2},\\
&\chi_{eff}=\frac{1}{2}\left(\sqrt{1-4\nu}\chi_1+\frac{m_1\chi_1-m_2\chi_2}{m_1+m_2}\right),
\end{align}
where $(m_1,m_2)$ are the masses and $(\chi_1,\chi_2)$ are the spin parameters of the original  BHs. The energy radiation efficiency $\epsilon_{rd}$ appeared in Eq. (\ref{eq4}) is related to the amplitude $\mathcal{A}_{lm}$ by \cite{Cardoso}
\begin{equation}
\mathcal{A}_{lm}=\sqrt{\frac{32Q_{lm}\epsilon_{rd}}{M f_{lm}(1+4Q_{lm}^2)}}.\label{eq5}
\end{equation}
We have omitted the overtone index $n$ in our expressions because only $n=0$ modes are considered here. The SNR obtained by  Eq. (\ref{eq3}) is denoted by $\rho_{\mathrm{int}}$,
\begin{gather}
\rho_{\mathrm{int}}^2=4\int _{f_{low}}^{f_{high}}\frac{|\tilde{h}_+|^2+|\tilde{h}_{\times}|^2}{S_n(f)}df,
\end{gather}
where $f_{low}$ is taken to be the half of the $(2,1)$ mode oscillation frequency and $f_{high}$ is taken to be two times of the $(4,4)$ mode frequency, and we also take the angle average $<|S_{lm}|^2>=1/4\pi$ for all modes following the average made in Ref. \cite{Cardoso}. Now we have calculated $\rho_{FH}$ by adopting the approximate formula Eq. (\ref{eq4}) and $\rho_{\mathrm{int}}$ obtained from the direct integral (A5) for comparison. In Fig. \ref{fig9} we show the behavior of the ratio $\rho_{\mathrm{int}}/\rho_{FH}$ with the change of the black hole redshifted mass $M_z$ for four different single QNMs. This figure shows that the value of $\rho_{\mathrm{int}}$ is close to $\rho_{FH}$, which  suggests that the approximate formula Eq. (\ref{eq4}) we have applied to calculate SNR for TianQin and TaiJi is feasible  with acceptable errors in a single-mode wave detection.

\bibliographystyle{JHEP}

\bibliography{BHSNR}

\end{document}